\documentclass[11pt,a4paper]{ivoa}
\newif\iftth
\iftth
\newcommand{\ivoaDocversion}{1.0}
\newcommand{\ivoaDocdate}{2014-12-04}
\newcommand{\ivoaDocdatecode}{20141204}
\newcommand{\ivoaDoctype}{REC}
\newcommand{\ivoaDocname}{RegTAP}

\input tthntbib.sty

\begin{html}
<meta http-equiv="Content-type" content="text/html;charset=UTF-8"/>
\end{html}

\definecolor{ivoacolor}{rgb}{0.0,0.318,0.612}

\renewcommand{\author}[2][0]{\def\@tmp{#1}
  \if 0\@tmp
	{\begin{html}<li class="author">#2</li>\end{html}}\else
	{\begin{html}<li class="author"><a href="#1">#2</a></li>\end{html}}\fi}
\renewcommand{\previousversion}[2][0]{\def\@tmp{#1}
  \if 0\@tmp
	{\begin{html}<li class="previousversion">#2</li>\end{html}}\else
	{\begin{html}<li class="previousversion">
	  <a href="#1">#2</a></li>\end{html}}\fi}
\renewcommand{\ivoagroup}[1]
  {\begin{html}<dd id="ivoagroup">#1</dd>\end{html}}
\renewcommand{\editor}[2][0]{\def\@tmp{#1}
  \if 0\@tmp
        {\begin{html}<li class="editor">#2</li>\end{html}}\else
        {\begin{html}<li class="editor"><a href="#1">#2</a></li>\end{html}}\fi}

\newcommand{\includeMeta}{%
   \ivoaDocversion\ivoaDoctype\ivoaDocname\ivoaDocdate}

\def\SVN$#1: #2 ${%
	#2}

\newenvironment{abstract}{%
  \includeMeta
  \begin{html}
    </div> <!-- titlepage -->
    <div id="abstract"><h2>Abstract</h2>
  \end{html}
  }{%
    \ivoaDoctype
    \tableofcontents
  }

\newcommand{\lstloadlanguages}[1]{}
\newcommand{\lstset}[1]{}

\newenvironment{lstlisting}[1][plain]
  {\tthverbatim{lstlisting}}
  {}

\newenvironment{bigdescription}{%
    \begin{html}<div class="bigdescription">\end{html}
    \begin{description}
  }{\end{description}\begin{html}</div>\end{html}}

\newcommand{\specialterm}[2]{%
  \begin{html}<span class="#1">\end{html}#2\begin{html}</span>\end{html}}

\newcommand{\vorent}[1]{\specialterm{vorent}{#1}}

\newcommand{\sptablerule}{}

\def\dquote{"}

\newenvironment{inlinetable}{}{}

\newcommand{\harvarditem}[4][0]{%
  
  \if 0#1 \item[#2 (#3)]
  \else \item[#1 (#3)]\fi}
\newcommand{\harvardurl}[1]{\url{#1}}

\def\AtBeginDocument#1{\relax}
\def\pgfsyspdfmark#1#2#3{\relax}

\newbox\voidb@x
\def\@m{\relax}

\begin{html}
  <div id="titlepage">
\end{html}

\fi

\usepackage[utf8]{inputenc}
\usepackage{longtable}
\usepackage{listings}
\definecolor{rtcolor}{rgb}{0.15,0.4,0.3}
\definecolor{tapcolor}{rgb}{0.4,0.1,0.1}

\newcommand{\rtent}[1]{\texttt{\color{rtcolor} #1}}
\newcommand{\tapent}[1]{\texttt{\color{tapcolor} #1}}

\ivoagroup{Registry}

\author[http://www.ivoa.net/cgi-bin/twiki/bin/view/IVOA/MarkusDemleitner]{Markus Demleitner}
\author[http://www.ivoa.net/cgi-bin/twiki/bin/view/IVOA/PaulHarrison]{Paul Harrison}
\author[http://www.ivoa.net/cgi-bin/twiki/bin/view/IVOA/MarcoMolinaro]{Marco Molinaro}
\author[http://www.ivoa.net/cgi-bin/twiki/bin/view/IVOA/GretchenGreene]{Gretchen Greene}
\author[http://www.ivoa.net/cgi-bin/twiki/bin/view/IVOA/TheresaDower]{Theresa Dower}
\author[http://wiki.ivoa.net/twiki/bin/view/IVOA/MenelaosPerdikeas]{Menelaos Perdikeas}

\editor{Markus Demleitner}

\previousversion[http://www.ivoa.net/documents/RegTAP/20141030]{http://www.ivoa.net/documents/RegTAP/20141030}
\previousversion[http://www.ivoa.net/documents/RegTAP/20140227]{http://www.ivoa.net/documents/RegTAP/20140227}
\previousversion[http://www.ivoa.net/documents/RegTAP/20140627]{http://www.ivoa.net/documents/RegTAP/20140627}
\previousversion[http://www.ivoa.net/documents/RegTAP/20140227]{http://www.ivoa.net/documents/RegTAP/20140227}
\previousversion[http://www.ivoa.net/documents/RegTAP/20131203]{http://www.ivoa.net/documents/RegTAP/20131203}
\previousversion[http://www.ivoa.net/documents/RegTAP/20130909]{http://www.ivoa.net/documents/RegTAP/20130909}
\previousversion[http://www.ivoa.net/documents/RegTAP/20130411]{http://www.ivoa.net/documents/RegTAP/20130411}
\previousversion{Internal Working Draft 2013-03-05 (Volute
rev. 2011)}
\previousversion{ Internal Working Draft 2012-11-12 (Volute
rev. 1864)}

\title{IVOA Registry Relational Schema}

\begin{document}

\begin{abstract}
Registries provide a mechanism with which VO applications can
discover and select resources -- first and foremost data and
services -- that are relevant for a particular scientific problem.
This specification defines  an interface for searching this resource
metadata based on the IVOA's TAP protocol.  It specifies a set of tables
that comprise a useful subset of the information contained in the
registry records, as well as the table's data content in terms of the
XML VOResource data model.  The general design of the system is geared
towards allowing easy authoring of queries.
\end{abstract}



\section{Introduction}

\label{intro}

In the Virtual Observatory (VO), registries provide a means for
discovering useful resources, i.e., data and services.  Individual
publishers offer the descriptions for their resources (``resource
records'') in publishing registries.  At the time of writing, there are
roughly 14000 such resource records active within the VO, originating
from about 20 publishing registries.

The protocol spoken by these
publishing registries, OAI-PMH, only allows restricting queries by
modification date and identifier and is hence not suitable for data discovery.
Even if it were, data discovery would at least be fairly time consuming if
each client had to query dozens or, potentially, hundreds of
publishing registries.

To enable efficient data discovery nevertheless, there are services harvesting the
resource records from the publishing registries and offering rich query
languages.
The IVOA Registry
Interfaces specification \citep{std:RI1} defined, among other aspects of
the VO registry system, such an interface
using SOAP and an early draft of an XML-based query language.

This document provides an update to the query (``searchable registry'') part
of that specifiation (chapter 2), aimed towards
usage with current VO standards, in particular TAP \citep{std:TAP}
and ADQL \citep{std:ADQL}.  It follows the model of ObsCore
\citep{std:OBSCORE} of defining a representation of a data model
within a relational database.  In this case, the data model is a
simplification of the VO's resource metadata interchange representation,
the VOResource XML format \citep{std:VOR}.  The simplification
yields 13 tables.  For each table, TAP\_SCHEMA metadata is given together
with rules for how to fill these tables from VOResource-serialized
metadata records as well as conditions on foreign keys and
recommendations on indexes.

The resulting set of tables has a modest size by today's standards,
but is still non-trivial.  The largest table, \rtent{table\_column},
has about half a million rows at the time of writing.

The architecture laid out here allows client applications to perform ``canned''
queries on behalf of their users as well as complex queries formulated
directly by advanced users, using the same TAP clients they employ to
query astronomical data servers.


\subsection{Terminology}

\label{terms}

The set of tables and their metadata specified here, together with
the mapping from VOResource (``ingestion rules'') is collectively
called ``relational registry schema'' or ``relational registry'' for
short.

The specificiation additionally talks about how to embed these into TAP
services, gives additional user defined functions, talks about
discovering compliant services, etc.  Since all this is tightly coupled
to the ``relational registry'' as defined above, we do not 
introduce a new term for it.  Hence, the entire standard is now known as
``IVOA registry relational schema''.

Historically, we intended to follow the ObsCore/ObsTAP model and
talked about RegTAP.  As changing this acronym is technically painful
(e.g., identifiers and URLs would need to be adapted), we kept it even after
the distinction between the schema and its mapping on the one hand and
its combination with a TAP service on the other went away.  This
means that the official acronym for ``IVOA registry relational schema'' is
RegTAP.  This aesthetic defect seems preferable to causing actual
incompatibilities.



\subsection{The Relational Registry within the VO Architecture}

\label{rolewithinivoa}

\begin{figure}[thm]
\begin{center}
\includegraphics[width=0.9\textwidth]{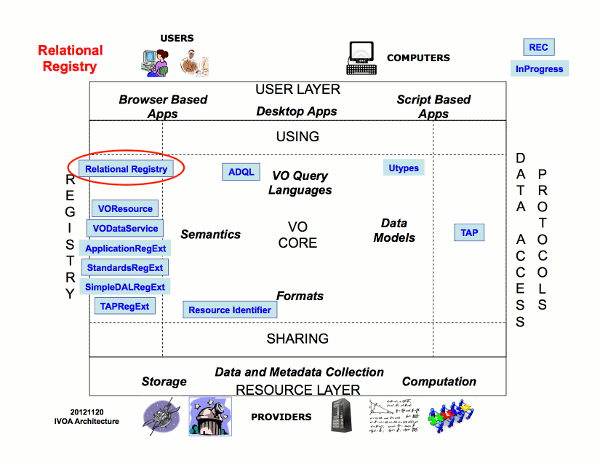}
\end{center}

\caption{IVOA Architecture
diagram with the IVOA Registry Relational Specification (tagged with
``Relational Registry'') and the related standards marked up.}
\end{figure}

This specification directly relates to other VO standards in the
following ways:

\begin{description}
\item[VOResource, v1.03 \citep{std:VOR}] VOResource sets the foundation for a formal definition of the data
model for resource records via its schema definition.  This document
refers to concepts laid down there via xpaths \citep{std:XPATH}.
\item[VODataService, v1.1 \citep{std:VODS11}] The VODataService standard 
describes several concepts and resource types extending 
VOResource's data model, including
tablesets, data services and data
collections.  These concepts and types are reflected in the database
schema.  Again xpaths link this specification and VODataService.
\item[Other Registry Extensions]Registry extensions are VO standards
defining how particular resources (e.g., Standards) or capabilities
(e.g., IVOA defined interfaces) are described.  Most aspects
introduced by them are reflected in the \rtent{res\_detail} table using
xpaths into the registry documents.
This document should not in general need updates
for registry extension updates.  For completeness, we note the
versions current as of this specification: SimpleDALRegExt 1.0
\citep{std:DALREGEXT},
StandardsRegExt 1.0 \citep{std:STDREGEXT}, TAPRegExt 1.0
\citep{std:TAPREGEXT}, Registry Interfaces 1.0 \citep{std:RI1}.
\item[TAP, v1.0 \citep{std:TAP}]The queries against the schema defined in the present document, and the results of
these queries, will usually be transported using the Table Access
Protocol TAP.  It also allows discovering
local additions to the registry relations via TAP's metadata publishing
mechanisms.
\item[IVOA Identifiers, v1.12 \citep{std:VOID}]IVOA identifiers are
essentially the primary keys within the VO
registry; as such, they are actual primary keys of the central table of
the relational registry. Also, the notion of an authority as laid down
in IVOA Identifiers plays an important role as publishing registries can
be viewed as a realization of a set of authorities.

\end{description}

This standard also relates to other IVOA standards:

\begin{description}
\item[ADQL, v2.0 \citep{std:ADQL}]The rules for ingestion are designed to allow
easy queries given the constraints of ADQL 2.0.  Also,
we give four functions that extend ADQL using the
language's built-in facility for user-defined functions.
\end{description}




\section{Design Considerations}

\label{design}

In the design of the tables, the goal has been to preserve as much of
VOResource and its extensions, including the element names, as
possible.

An overriding consideration has been, however, to make natural joins
between the tables behave usefully, i.e., to actually combine rows
relevant to the same entity (resource, table, capability, etc.).
To disambiguate column names that name the same concept on different
entities (name, description, etc.) and would therefore interfere with
the natural join, a shortened tag for the source object
is prepended to the name.  Thus, a \vorent{description} element within
a resource ends up in a column named
\rtent{res\_description}, whereas the same element from a
\vorent{capability} becomes \rtent{cap\_description}.

We further renamed some columns and most tables 
with respect to their VOResource
counterparts to avoid clashes with reserved words in popular database
management systems.  The alternatives would have been to either recommend
quoting them or burden ADQL translation layers with the task of
automatically converting them to delimited identifiers.  Both
alternatives seemed more confusing and less robust than the renaming
proposed here.

Furthermore, camel-case identifiers have been converted to
underscore-separated ones (thus, \vorent{standardId}  becomes
\rtent{standard\_id}) to have all-lowercase column names; this saves
potential headache if users choose to reference the columns using SQL
delimited identifiers.  Dashes in VOResource attribute names are
converted to underscores, too, with the exception of
\vorent{ivo-id}, which is just rendered \rtent{ivoid}.

Another design goal of this specification has been that different registries
operating on the same set of registry records will return identical responses
for most queries; hence, we try to avoid relying on features left not
defined by ADQL (e.g., the case sensitivity of string matches).  However,
with a view to non-uniform support for information retrieval-type
queries in database systems, the \rtent{ivo\_hasword} user defined
function is not fully specified here; queries employing it may yield
different results on different implementations, even if they operate on
the same set of resource records.



\section{Primary Keys}

\label{primarykeys}

The primary key in the Registry as an abstract concept is a resource
record's IVORN.  Hence, for all tables having primary keys at all, the
\rtent{ivoid} column is part of its primary key.  This
specification does not require implementations to actually declare
primary keys in the underlying database, and no aspect of user-visible
behavior depends on such explicit declarations; in particular, this
specification makes no requirements on the contents of
\tapent{tap\_schema.keys}.

We nevertheless make recommendations on explicit primary keys, as
we expect definitions according to our recommendations will enhance
robustness of services.

In several RegTAP tables -- \rtent{capability},
\rtent{res\_schema}, \rtent{res\_table}, and
\rtent{interface} -- artificial primary keys are necessary, as
in VOResource XML sibling elements are not otherwise distinguished.  To
allow such artificial primary keys, a column is added to each table, the
name of which ends in \texttt{\_index} (\rtent{cap\_index},
\rtent{schema\_index}, \rtent{table\_index}, and
\rtent{intf\_index}).

The type and content of these \texttt{X\_index} columns is
implementation-defined, and clients must not make assumptions on their
content except that the pair \rtent{ivoid}, \texttt{X\_index} is a primary
key for the relation (plus, of course, that references from other tables
correctly resolve).  In the tables of columns given below, the
\texttt{X\_index} columns have ``(key)'' given for type.  Implementors
obviously have to insert whatever ADQL type is appropriate for their
choice or \texttt{X\_index} implementation.

Obvious implementations for \texttt{X\_index} include having
\texttt{X\_index} enumerate the sibling elements or using some sort
of UUID.



\section{Notes on string handling}

\label{stringnorm}

In the interest of consistent behavior between different RegTAP
implementations regardless of their technology choices, this section
establishes some rules on the treatment of strings -- both those
obtained from attributes and those obtained from element
content -- during ingestion from VOResource XML to database
tables.


\subsection{Whitespace normalization}

\label{whitenorm}

Most string-valued items in VOResource and extensions are of type
\texttt{xs:to\-ken}, with the clear intent that whitespace in them is
to be normalized in the sense of XML schema.  For the few exceptions
that actually are directly derived from xs:string (e.g.,
\vorent{vstd:EndorsedVersion}, \vorent{vs:Waveband}) it does not
appear that the intent regarding whitespace is different.

In order to provide reliable querying and simple rules for ingestors
even when these do not employ schema-aware XML parsers, this standard
requires that during ingestion, leading and trailing whitespace MUST be
removed from all strings; in particular, there are no strings consisting
exclusively of whitespace in RegTAP.  The treatment of internal
whitespace is implementation-defined. This reflects the expectation
that, whereever multi-word items are queried, whitespace-ignoring
constraints will be used (e.g., LIKE-based regular expressions or the
\rtent{ivo\_hasword} user defined function defined below).



\subsection{NULL/Empty string normalization}

\label{nullnorm}

While empty strings and NULL values are not usually well
distinguished in VO practice -- as reflected in the conventional
TABLEDATA and BINARY serializations of VOTable -- , the distinction
must be strictly maintained in the database tables to ensure
reproduceable queries across different RegTAP implementations.

Ingestors therefore MUST turn empty strings (which, by section \ref{whitenorm}, include strings consisting of whitespace
only in VOResource's XML serialization) into NULL values in the
database.  Clients expressing constraints on the presence (or absence)
of some information must therefore do so using SQL's \texttt{IS NOT NULL}
(or \texttt{IS NULL}) operators.



\subsection{Case normalization}

\label{casenorm}

ADQL has no operators for case-insensitive matching of strings.  To
allow for robust and straightforward queries nevertheless, most columns
containing values not usually intended for display are required to be
converted to lower case; in the table descriptions below, there are
explicit requirements on case normalization near the end of each
section.  This is particularly important when the entities to be
compared are defined to be case-insensitive (e.g., UCDs, IVORNs).
Client software that can inspect user-provided arguments (e.g., when
filling template queries) should also convert the respective fields to
lower case.

This conversion MUST cover all ASCII letters, i.e., A through Z.  
The conversion SHOULD take place according to
algorithm R2 in section 3.13, ``Default Case Algorithms'' of the Unicode
Standard
\citep{std:UNICODE}.  In practice, non-ASCII characters are not expected
to occur in columns for which lowercasing is required.

Analogously, case-insensitive comparisons as required by some of the
user-defined functions for the relational registry MUST compare
the ASCII letters without regard for case.  They SHOULD compare according
to D144 in the Unicode Standard.



\subsection{Non-ASCII characters}

\label{utfreq}

Neither TAP nor ADQL mention non-ASCII in service parameters -- in
particular the queries -- or returned values.  For RegTAP, that is
unfortunate, as several columns will contain relevant non-ASCII
characters.  Columns for which extra care is necessary include all
descriptions, \rtent{res\_title} and \rtent{creator\_seq} in
\rtent{rr.resource}, as well as \rtent{role\_name} and
\rtent{street\_address} in \rtent{rr.res\_role}.

RegTAP implementations SHOULD be able to faithfully represent all
characters defined in the latest version of the Unicode standard 
\citep{std:UNICODE} at
any given time and allow querying using them (having support for UTF-8
in the database should cover this requirement) for at least the fields
mentioned above.

On VOResource ingestion, non-ASCII characters that a service cannot
faithfully store MUST be replaced by a question mark character (``?'').

RegTAP services MUST interpret incoming ADQL as encoded in UTF-8,
again replacing unsupported characters with question marks.

We leave character replacement on result generation unspecified, as
best-effort representations (e.g., ``Angstrom'' instead of ``Ångström'')
should not impact interoperability but significantly improve user
experience over consistent downgrading.  In VOTable output,
implementations SHOULD support full Unicode in at least the fields
enumerated above.  Clients are advised to retrieve results in VOTable or
other encoding-aware formats.

Note that with VOTable 1.3, non-ASCII in char-typed fields, while
supported by most clients in TABLEDATA serialization, is technically
illegal; it is essentially undefined in other serializations.  To
produce standards-compliant VOTables, columns containing non-ASCII must
be of type unicodeChar.  We expect that future versions of VOTable will
change the definitions of char and unicodeChar to better match modern
standards and requirements.  RegTAP implementors are encouraged to take
these up.




\section{QNames in VOResource attributes}

\label{qnameatts}

VOResource and its extensions make use of XML QNames in attribute
values, most prominently in \texttt{xsi:type}.  The standard
representation of these QNames in XML instance documents makes use of an
abbreviated notation employing prefixes declared using the xmlns mechanism
as discussed in \citet{std:XMLNS}.  Within an ADQL-exposed database, no
standard mechanism exists that could provide a similar mapping of URLs
and abbreviations.  The correct way to handle this problem would thus be
to have full QNames in the database (e.g.,
\texttt{{http://www.ivoa.net/xml/ConeSearch/v1.0}ConeSearch} for the
canonical \vorent{cs:ConeSearch}).  This, of course, would make for
excessively tedious and error-prone querying.

For various reasons, VOResource authors have always been encouraged
to use a set of ``standard'' prefixes.  This allows an easy and, to users,
unsurprising exit from the problem of the missing xmlns declarations:
For the representation of QNames within the database, these recommended
prefixes are now mandatory. Future VOResource extensions define their
mandatory prefixes themselves.

Following the existing practice, minor version changes are not in
general reflected in the recommended prefixes -- e.g., both VODataService
1.0 and VODataService 1.1 use \texttt{vs:}.  For reference, here is
a table of XML namespaces and prefixes for namespaces relevant to this
specification:

\begin{inlinetable}
\begin{tabular}{ll}
cs&http://www.ivoa.net/xml/ConeSearch/v1.0\\
dc&http://purl.org/dc/elements/1.1/\\
oai&http://www.openarchives.org/OAI/2.0/\\
ri&http://www.ivoa.net/xml/RegistryInterface/v1.0\\
sia&http://www.ivoa.net/xml/SIA/v1.0\\
sia&http://www.ivoa.net/xml/SIA/v1.1\\
slap&http://www.ivoa.net/xml/SLAP/v1.0\\
ssap&http://www.ivoa.net/xml/SSA/v1.0\\
ssap&http://www.ivoa.net/xml/SSA/v1.1\\
tr&http://www.ivoa.net/xml/TAPRegExt/v1.0\\
vg&http://www.ivoa.net/xml/VORegistry/v1.0\\
vr&http://www.ivoa.net/xml/VOResource/v1.0\\
vs&http://www.ivoa.net/xml/VODataService/v1.0\\
vs&http://www.ivoa.net/xml/VODataService/v1.1\\
vstd&http://www.ivoa.net/xml/StandardsRegExt/v1.0\\
xsi&http://www.w3.org/2001/XMLSchema-instance\\

\end{tabular}
\end{inlinetable}



\section{Xpaths}

\label{vorutypes}

This specification piggybacks on top of the well-established
VOResource standard.  This means that it does not define a full data model,
but rather something like a reasonably query-friendly view of a partial
representation of one.  The link between the actual data model, i.e.,
VOResource and its extensions as defined by the XML Schema documents, and
the fields within this database schema, is provided by
xpaths, which are here slightly abbreviated for both brevity and
generality.

All xpaths given in this specifiation are assumed to be relative to
the enclosing \vorent{vr:Resource} element; these are called
``resource xpaths'' in the following.  If resource xpaths are to be
applied to an OAI-PMH response, the Xpath expression
\texttt{*/*/*/oai:metadata/ri:Resource} must be prepended to it,
with the canonical prefixes from section \ref{qnameatts} implied.  The resource xpaths themselves
largely do not need explicit namespaces since VOResource elements are by
default unqualified.  Elements and attributes from non-VOResource
schemata have the canonical namespace prefixes, which in this
specification only applies to several \texttt{xsi:type} attribute
names.

Some tables draw data from several different VOResource elements.
For those, we have introduced an extended syntax with additional
metacharacters \verb$($, \verb$)$, and \verb$|$, 
where the vertical bar denotes an
alternative and the parentheses grouping.  For instance, our notation
\texttt{/(tableset/schema/|)table/} corresponds to the two xpaths
\texttt{/table} and \texttt{/tableset/schema/table}.

Within the Virtual Observatory, the link between data models and
concrete data representations is usually made using utypes.
Since VOResource is directly modelled
in XML Schema, the choice of XPath as the bridging formalism is 
compelling, though, and utypes themselves are not necessary for the
operation of a TAP service containing the relational registry.
TAP, however, offers fields for utypes in its TAP\_SCHEMA.  Since they
are not otherwise required, this specification takes the liberty of
using them to denote the xpaths.

In the metadata for tables and columns below, the utypes given are
obtained from the xpaths by simply prepending them with
\texttt{xpath:}.  To avoid repetition, we allow relative xpaths:
when the xpath in a column utype does not start with a slash, it is
understood that it must be concatenated with the table utype to obtain
the full xpath.

For illustration, if a table has a utype of
$$\texttt{xpath:/capability/interface/}$$ and a column within this table
has a utype of $$\texttt{xpath:accessURL/@use},$$ the resulting resource
xpath would come out to be
$$\texttt{/capability/interface/accessURL/@use};$$ to match this in an
OAI-PMH response, the XPath would be
$$\texttt{\small
*/*/*/oai:metadata/ri:Resource/capability/interface/accessURL/@use}.$$

While clients MUST NOT rely on these utypes in either
\tapent{TAP\_SCHEMA} or the
metadata delivered with TAP replies, service operators SHOULD provide them, in
particular when there are local extensions to the relational registry in their
services.  Giving xpaths for extra columns and tables helps human
interpretation of them at least when the defining schema files are
available.

Resource xpaths are also used in the \rtent{res\_details} table (section
\ref{table_res_detail}).  These are normal xpaths
(although again understood relative to the enclosing Resource element),
which, in particular, means that they are case sensitive.  On the other
hand, to clients they are simply opaque strings, i.e., clients cannot
just search for any xpaths into VOResource within \rtent{res\_details}.



\section{Discovering Relational Registries}

\label{registration}

The relational registry can be part of any TAP service.  The presence
of the tables discussed here is indicated by declaring support for the
data model \texttt{Registry 1.0} with the IVORN
$$\texttt{ivo://ivoa.net/std/RegTAP\#1.0}$$ in the service's
capabilities (cf.~\citet{std:TAPREGEXT}).  Technically, this
entails adding

\begin{verbatim}
<dataModel ivo-id="ivo://ivoa.net/std/RegTAP#1.0"
  >Registry 1.0</dataModel>
\end{verbatim}

as a child of the capability element with the type
\vorent{tr:TableAccess}.

A client that knows the access URL of one TAP service containing 
a relational
registry can thus discover all other services exposing one. The \href{#example-findregtap}{``Find all TAP endpoints offering the
relational registry''} example in section
\ref{sample_queries} shows a query that does
this.

It is recommended to additionally register a relational registry as a
VODataService data collection and connect this and the TAP services
with a pair of service-for/served-by relations.  This allows giving more
metadata on the data content like, for example, the frequency of
harvesting.

Services implementing this data model that do not (strive to) offer
the full data content of the VO registry (like domain-specific
registries or experimental systems) MUST NOT declare the above data
model in order to not invite clients expecting the VO registry to send
queries to it.



\section{VOResource Tables}

\label{vortables}

In the following table descriptions, the names of tables
(cf.~Table \ref{table:dm}) and columns
are normative and MUST be used as given, and all-lowercase.  On the
values in the \tapent{utype} columns within \tapent{TAP\_SCHEMA},
see section \ref{vorutypes}.  All columns defined in
this document MUST have a 1 in the \tapent{std} column of the
\tapent{TAP\_SCHEMA.table\_columns} table.  Unless otherwise
specified, all values of ucd and unit in
\tapent{TAP\_SCHEMA.table\_columns} are NULL for columns defined here.
Descriptions are not normative (as given, they usually are taken from
the schema files of VOResource and its extensions with slight
redaction).  Registry operators MAY provide additional columns in their
tables, but they MUST provide all columns given in this
specification.

All table descriptions start out with brief remarks on the
relationship of the table to the VOResource XML data model.  Then, the
columns are described in a selection of TAP\_SCHEMA metadata. For each
table, recommendations on explicit primary and foreign keys as well as
indexed columns are given, where it is understood that primary and
foreign keys are already indexed in order to allow efficient joins;
these parts are not normative, but operators should ensure decent
performance for queries assuming the presence of the given indexes and
relationships.  Finally, lowercasing requirements
(normative) are given.

\begin{figure}

\includegraphics[width=\textwidth]{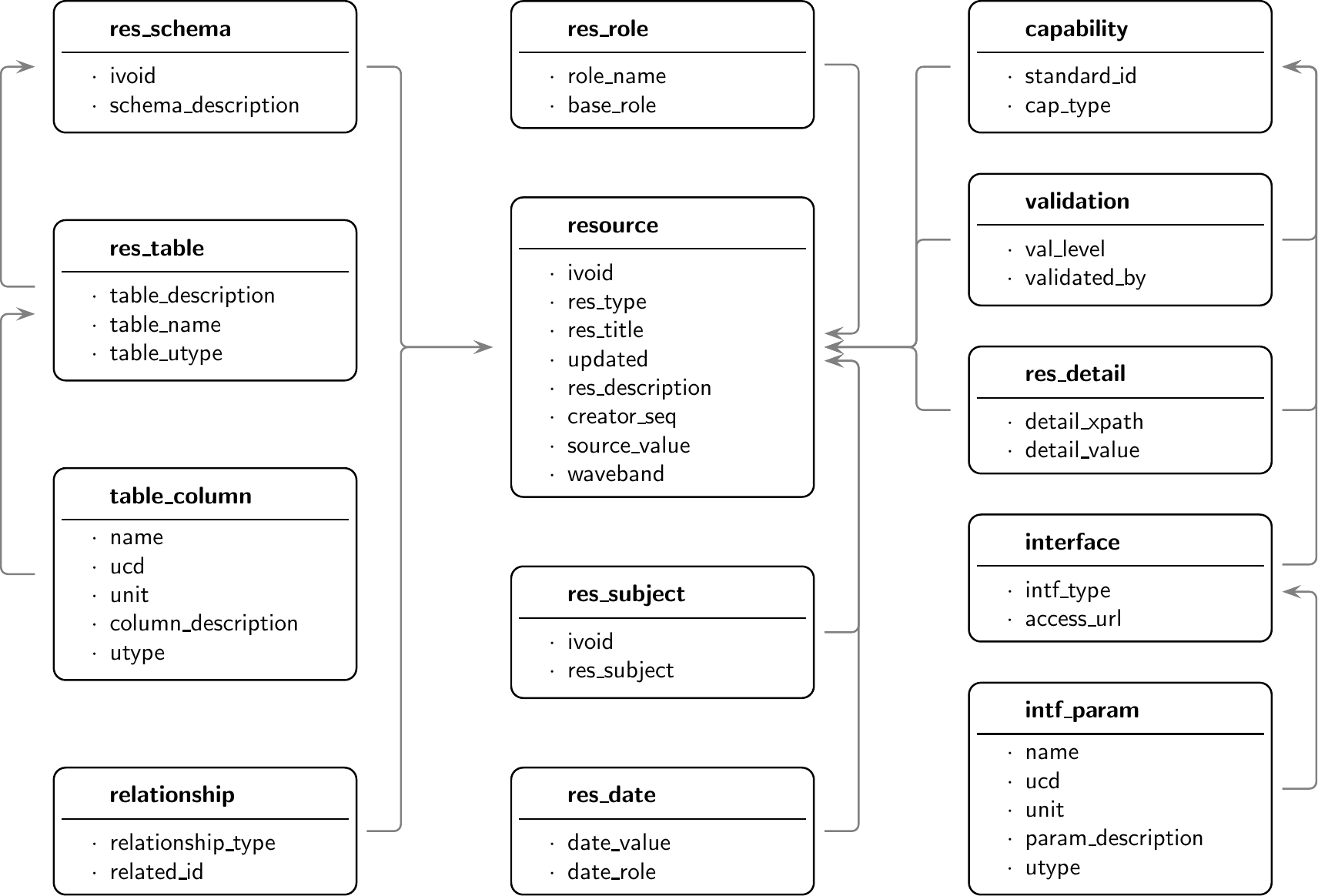}
\caption{A sketch of the
Relational Registry schema, adapted from \citet{regtap-adass}.  
Only the columns considered
most interesting for client use are shown.  Arrows indicate foreign
key-like relationships.}
\end{figure}


\begin{table}[t]
\small
\hbox to\hsize{\hss
\begin{tabular}{p{0.35\textwidth}p{0.64\textwidth}}
\sptablerule
\textbf{Name and UType}&\textbf{Description}\\
\sptablerule
rr.capability\hfil\break
\makebox[0pt][l]{\scriptsize\ttfamily xpath:/capability/}&
 Pieces of behaviour of a resource.\\
rr.interface\hfil\break
\makebox[0pt][l]{\scriptsize\ttfamily xpath:/capability/interface/}&
 Information on access modes of a capability.\\
rr.intf\_param\hfil\break
\makebox[0pt][l]{\scriptsize\ttfamily xpath:/capability/interface/param/}&
 Input parameters for services.\\
rr.relationship\hfil\break
\makebox[0pt][l]{\scriptsize\ttfamily xpath:/content/relationship/}&
 Relationships between resources, e.g., mirroring, derivation, but
also providing access to data within a resource.\\
rr.res\_date\hfil\break
\makebox[0pt][l]{\scriptsize\ttfamily xpath:/curation/}&
 A date associated with an event in the life cycle of the resource.
This could be creation or update. The role column can be used to
clarify.\\
rr.res\_detail\hfil\break
\makebox[0pt][l]{\scriptsize\ttfamily }&
 XPath-value pairs for members of resource or capability and their
derivations that are less used and/or from VOResource extensions. The
pairs refer to a resource if cap\_index is NULL, to the referenced
capability otherwise.\\
rr.res\_role\hfil\break
\makebox[0pt][l]{\scriptsize\ttfamily }&
 Entities, i.e., persons or organizations, operating on resources:
creators, contacts, publishers, contributors.\\
rr.res\_schema\hfil\break
\makebox[0pt][l]{\scriptsize\ttfamily xpath:/tableset/schema/}&
 Sets of tables related to resources.\\
rr.res\_subject\hfil\break
\makebox[0pt][l]{\scriptsize\ttfamily xpath:/content/}&
 Topics, object types, or other descriptive keywords about the
resource.\\
rr.res\_table\hfil\break
\makebox[0pt][l]{\scriptsize\ttfamily xpath:/(tableset/schema/|)table/}&
 (Relational) tables that are part of schemata or resources.\\
rr.resource\hfil\break
\makebox[0pt][l]{\scriptsize\ttfamily xpath:/}&
 The resources, i.e., services, data collections, organizations, etc.,
present in this registry.\\
rr.table\_column\hfil\break
\makebox[0pt][l]{\scriptsize\ttfamily xpath:/(tableset/schema/|)/table/column/}&
 Metadata on columns of a resource's tables.\\
rr.validation\hfil\break
\makebox[0pt][l]{\scriptsize\ttfamily xpath:/(capability|)}&
Validation levels for resources and capabilities.\\

\sptablerule
\end{tabular}\hss}
\caption{The tables making up the TAP data model \texttt{Registry 1.0}}
\label{table:dm}
\end{table}



\subsection{The resource Table}

\label{table_resource}

The \rtent{rr.resource} table contains most atomic members of
\rtent{vr:Resource} that have a 1:1 relationship to the resource
itself.  Members of derived types are, in general, handled through 
the \rtent{res\_detail}
table even if 1:1 (see \ref{table_res_detail}).  The
\rtent{content\_level}, \rtent{content\_type}, \rtent{waveband}, 
and \rtent{rights} members are 1:n but still appear
here.  If there are multiple values, they are concatenated with hash
characters (\#).  Use the \rtent{ivo\_hashlist\_has} ADQL extension
function to check for the presence of a single value.  This convention
saves on tables while not complicating common queries significantly.

A local addition is the \rtent{creator\_seq} column.  It contains
all content of the \vorent{name} elements below a resource element
\vorent{curation} child's \vorent{creator} children, concatenated with a
sequence of semicolon and blank characters (``\mbox{\texttt{; }}''). The
individual parts must be concatenated preserving the sequence of the XML
elements.  The resulting string is primarily intended for display
purposes (``author list'') and is hence not case-normalized.  It was
added since the equivalent of an author list is expected to be a
metadatum that is displayed fairly frequently, but also since the
sequence of author names is generally considered significant.  The
\rtent{res\_role} table, on the other hand, does not allow recovering
the input sequence of the rows belonging to one resource.

The \rtent{res\_type} column reflects the lower-cased value of
the \vorent{ri:Resource} element's \texttt{xsi:type} attribute,
where the canonical prefixes are used.  While custom or experimental
VOResource extensions may lead to more or less arbitrary strings in that
column, VOResource and its IVOA-recommended extensions at the time of
writing define the following values for \rtent{res\_type}:

\begin{description}
\item[vg:authority]A naming authority (these records allow resolving who is responsible
for IVORNs with a certain authority; cf. \citet{std:VOID}).
\item[vg:registry]A registry is considered a publishing registry if it contains a
capability element with \texttt{xsi:type="vg:Harvest"}.  Old,
RegistryInterface 1-compliant registries also use this type with a
capability of type \vorent{vg:Search}.  The relational registry as
specified here, while superceding these old \vorent{vg:Search}
capabilities, does \emph{not} use this type any more.  See section
\ref{registration} on how to locate services
supporting it.
\item[vr:organisation]The main purpose of an organisation as a registered resource is to
serve as a publisher of other resources.
\item[vr:resource]Any entity or component of a VO application that is describable and
identifiable by an IVOA identifier; while it is technically possible to
publish such records, the authors of such records should probably be
asked to use a more specific type.
\item[vr:service]A resource that can be invoked by a client to perform some action on
its behalf.
\item[vs:catalogservice]A service that interacts with one or more specified tables having
some coverage of the sky, time, and/or frequency.
\item[vs:dataservice]A service for accessing astronomical data; publishers choosing
this over \vorent{vs:catalogservice} probably intend to communicate
that there are no actual sky positions in the tables exposed.
\item[vs:datacollection]A schema as a logical grouping of data which, in general, is
composed of one or more accessible datasets.  Use the
\rtent{rr.relationship} table to find out services that allow
access to the data (the \vorent{served\_by} relation), and/or look for values for
\vorent{/accessURL} in
\ref{table_res_detail}.
\item[vstd:standard]A description of a standard specification.

\end{description}

The \vorent{status} attribute of \vorent{vr:Resource} is
considered an implementation detail of the XML serialization and is not
kept here.  Neither \vorent{inactive} nor \vorent{deleted}
records may be kept in the \rtent{resource} table.  Since all
other tables in the relational registry should keep a foreign key on the
\rtent{ivoid} column, this implies that only metadata on
\vorent{active} records 
is being kept in the relational registry. In other words, users can
expect a resource to exist and work if they find it in a relational 
registry.


\begin{inlinetable}
\small
\begin{tabular}{p{0.28\textwidth}p{0.2\textwidth}p{0.66\textwidth}}
\sptablerule
\multicolumn{3}{l}{\textit{Column names, utypes, ADQL types, and descriptions for the \rtent{rr.resource} table}}\\
\sptablerule
ivoid\hfil\break
\makebox[0pt][l]{\scriptsize\ttfamily xpath:identifier}&
\footnotesize VARCHAR(*)&
Unambiguous reference to the resource conforming to the IVOA standard for identifiers.\\
res\_type\hfil\break
\makebox[0pt][l]{\scriptsize\ttfamily xpath:@xsi:type}&
\footnotesize VARCHAR(*)&
Resource type (something like vs:datacollection, vs:catalogservice, etc).\\
created\hfil\break
\makebox[0pt][l]{\scriptsize\ttfamily xpath:@created}&
\footnotesize TIMESTAMP(1)&
The UTC date and time this resource metadata description was created. This timestamp must not be in the future. This time is not required to be accurate; it should be at least accurate to the day. Any insignificant time fields should be set to zero.\\
short\_name\hfil\break
\makebox[0pt][l]{\scriptsize\ttfamily xpath:shortName}&
\footnotesize VARCHAR(*)&
A short name or abbreviation given to something. This name will be used where brief annotations for the resource name are required. Applications may use it to refer to the resource in a compact display. One word or a few letters is recommended. No more than sixteen characters are allowed.\\
res\_title\hfil\break
\makebox[0pt][l]{\scriptsize\ttfamily xpath:title}&
\footnotesize VARCHAR(*)&
The full name given to the resource.\\
updated\hfil\break
\makebox[0pt][l]{\scriptsize\ttfamily xpath:@updated}&
\footnotesize TIMESTAMP(1)&
The UTC date this resource metadata description was last updated. This timestamp must not be in the future. This time is not required to be accurate; it should be at least accurate to the day. Any insignificant time fields should be set to zero.\\
content\_level\hfil\break
\makebox[0pt][l]{\scriptsize\ttfamily xpath:content/contentLevel}&
\footnotesize VARCHAR(*)&
A hash-separated list of content levels specifying the intended audience.\\
res\_description\hfil\break
\makebox[0pt][l]{\scriptsize\ttfamily xpath:content/description}&
\footnotesize VARCHAR(*)&
An account of the nature of the resource.\\
reference\_url\hfil\break
\makebox[0pt][l]{\scriptsize\ttfamily xpath:content/referenceURL}&
\footnotesize VARCHAR(*)&
URL pointing to a human-readable document describing this resource.\\
creator\_seq\hfil\break
\makebox[0pt][l]{\scriptsize\ttfamily xpath:curation/creator/name}&
\footnotesize VARCHAR(*)&
The creator(s) of the resource in the order given by the resource record author, separated by semicolons.\\
content\_type\hfil\break
\makebox[0pt][l]{\scriptsize\ttfamily xpath:content/type}&
\footnotesize VARCHAR(*)&
A hash-separated list of natures or genres of the content of the resource.\\
source\_format\hfil\break
\makebox[0pt][l]{\scriptsize\ttfamily xpath:content/source/@format}&
\footnotesize VARCHAR(*)&
The format of source\_value. Recognized values include "bibcode", referring to a standard astronomical bibcode (http://cdsweb.u-strasbg.fr/simbad/refcode.html).\\
source\_value\hfil\break
\makebox[0pt][l]{\scriptsize\ttfamily xpath:content/source}&
\footnotesize VARCHAR(*)&
A bibliographic reference from which the present resource is derived or extracted.\\
res\_version\hfil\break
\makebox[0pt][l]{\scriptsize\ttfamily xpath:curation/version}&
\footnotesize VARCHAR(*)&
Label associated with creation or availablilty of a version of a resource.\\
region\_of\_regard\hfil\break
\makebox[0pt][l]{\scriptsize\ttfamily xpath:coverage/regionOfRegard}&
\footnotesize REAL(1)&
A single numeric value representing the angle, given in decimal degrees, by which a positional query against this resource should be "blurred" in order to get an appropriate match.\\
waveband\hfil\break
\makebox[0pt][l]{\scriptsize\ttfamily xpath:coverage/waveband}&
\footnotesize VARCHAR(*)&
A hash-separated list of regions of the electro-magnetic spectrum that the resource's spectral coverage overlaps with.\\
rights\hfil\break
\makebox[0pt][l]{\scriptsize\ttfamily xpath:rights}&
\footnotesize VARCHAR(*)&
Information about rights held in and over the resource (multiple values are separated by hashes).\\

\sptablerule
\end{tabular}
\end{inlinetable}


This table should have the \rtent{ivoid} column explicitly set
as its primary key.

The following columns MUST be lowercased during ingestion:
\rtent{ivoid}, \rtent{res\_type}, \rtent{content\_level},
\rtent{content\_type}, \rtent{source\_format},
\rtent{waveband}.
Clients are advised to query the \rtent{res\_description} and
\rtent{res\_title}  columns
using the the \rtent{ivo\_hasword} function, and to use
\rtent{ivo\_hashlist\_has} on \rtent{content\_level},
\rtent{content\_type},
\rtent{waveband}, and \rtent{rights}.

The row for \rtent{region\_of\_regard} in  
\tapent{TAP\_SCHEMA.columns} MUST have \texttt{deg} in its
\tapent{unit} column.



\subsection{The res\_role Table}

\label{table_res_role}

This table subsumes the contact, publisher, contributor, 
and creator members of the
VOResource data model.  They have been combined into a single table to
reduce the total number of tables, and also in anticipation of a unified
data model for such entities in future versions of VOResource.

The actual role is given in the \rtent{base\_role} column, which
can be one of \rtent{contact}, \rtent{publisher}, \rtent{contributor}, or
\rtent{creator}.  Depending on this value, here are the xpaths
for the table fields (we have abbreviated
\vorent{/curation/publisher}
as cp, \vorent{/curation/contact} as co, \vorent{/curation/creator} 
as cc,
and \vorent{/curation/contributor} as cb):

\vspace{5pt}
\hbox to\hsize{\hss\small
\noindent\begin{tabular}{lllll}
\sptablerule
\textbf{base\_role value}&
\textbf{contact}&
\textbf{publisher}&
\textbf{creator}&
\textbf{contributor}\\
\sptablerule
role\_name&co/name&cp&cc/name&cb\\
role\_ivoid&co/name/@ivo-id&cp/@ivo-id&cc/name/@ivo-id&cb/@ivo-id\\
address&co/address&N/A&N/A&N/A\\
email&co/email&N/A&N/A&N/A\\
telephone&co/telephone&N/A&N/A&N/A\\
logo&co/logo&N/A&cc/logo&N/A\\
\sptablerule
\end{tabular}\hss}
\vskip5pt

Not all columns are available for each role type in VOResource.  For
example, contacts have no logo, and creators no telephone members.  Unavailable
metadata (marked with N/A in the above table) MUST be represented with NULL
values in the corresponding columns.

Note that, due to current practice in the VO, it is not easy to
predict what \rtent{role\_name} will contain; it could be a single
name, where again the actual format is unpredictable
(full first name, initials in front or behind, or
even a project name), but it could as well be a full author list.  Thus,
when matching against \rtent{role\_name}, you will have to use
rather lenient regular expressions.  Changing this, admittedly
regrettable, situation would
probably require a change in the VOResource schema.


\begin{inlinetable}
\small
\begin{tabular}{p{0.28\textwidth}p{0.2\textwidth}p{0.66\textwidth}}
\sptablerule
\multicolumn{3}{l}{\textit{Column names, utypes, ADQL types, and descriptions for the \rtent{rr.res\_role} table}}\\
\sptablerule
ivoid\hfil\break
\makebox[0pt][l]{\scriptsize\ttfamily xpath:/identifier}&
\footnotesize VARCHAR(*)&
The parent resource.\\
role\_name\hfil\break
\makebox[0pt][l]{\scriptsize\ttfamily }&
\footnotesize VARCHAR(*)&
The real-world name or title of a person or organization.\\
role\_ivoid\hfil\break
\makebox[0pt][l]{\scriptsize\ttfamily }&
\footnotesize VARCHAR(*)&
An IVOA identifier of a person or organization.\\
street\_address\hfil\break
\makebox[0pt][l]{\scriptsize\ttfamily }&
\footnotesize VARCHAR(*)&
A mailing address for a person or organization.\\
email\hfil\break
\makebox[0pt][l]{\scriptsize\ttfamily }&
\footnotesize VARCHAR(*)&
An email address the entity can be reached at.\\
telephone\hfil\break
\makebox[0pt][l]{\scriptsize\ttfamily }&
\footnotesize VARCHAR(*)&
A telephone number the entity can be reached at.\\
logo\hfil\break
\makebox[0pt][l]{\scriptsize\ttfamily }&
\footnotesize VARCHAR(*)&
URL pointing to a graphical logo, which may be used to help identify the entity.\\
base\_role\hfil\break
\makebox[0pt][l]{\scriptsize\ttfamily }&
\footnotesize VARCHAR(*)&
The role played by this entity; this is one of contact, publisher, and creator.\\

\sptablerule
\end{tabular}
\end{inlinetable}


The \rtent{ivoid} column should be an explicit foreign key into
the \rtent{resource} table.  It is recommended to maintain indexes
on at least the \rtent{role\_name} column, ideally in a way that
supports regular expressions.

The following columns MUST be lowercased during ingestion:
\rtent{ivoid}, \rtent{role\_ivoid},
\rtent{base\_role}.
Clients are advised to query the remaining columns, in particular
\rtent{role\_name},
case-insensitively, e.g., using \rtent{ivo\_nocasematch}.



\subsection{The res\_subject Table}

\label{table_res_subject}

Since subject queries are expected to be frequent and perform relatively
complex checks (e.g., resulting from thesaurus queries in the clients), the
subjects are kept in a separate table rather than being hash-joined like other
string-like 1:n members of resource.


\begin{inlinetable}
\small
\begin{tabular}{p{0.28\textwidth}p{0.2\textwidth}p{0.66\textwidth}}
\sptablerule
\multicolumn{3}{l}{\textit{Column names, utypes, ADQL types, and descriptions for the \rtent{rr.res\_subject} table}}\\
\sptablerule
ivoid\hfil\break
\makebox[0pt][l]{\scriptsize\ttfamily xpath:/identifier}&
\footnotesize VARCHAR(*)&
The parent resource.\\
res\_subject\hfil\break
\makebox[0pt][l]{\scriptsize\ttfamily xpath:subject}&
\footnotesize VARCHAR(*)&
Topics, object types, or other descriptive keywords about the resource.\\

\sptablerule
\end{tabular}
\end{inlinetable}


The \rtent{ivoid}  column should be an explicit foreign key into
\rtent{resource}.  It is recommended to index the
\rtent{res\_subject} column, preferably in a way that allows to process
case-insensitive and pattern queries using the index.

The \rtent{ivoid} column MUST be lowercased during
ingestion.  Clients are advised to query the \rtent{res\_subject} column
case-insensitively, e.g., using \rtent{ivo\_nocasematch}.



\subsection{The capability Table}

\label{table_capability}

The capability table describes a resource's modes of interaction; it only
contains the members of the base type \vorent{vr:Capability}.
Members of derived types are kept in the \rtent{res\_detail} table
(see \ref{table_res_detail}).

The table has a
\rtent{cap\_index} to disambiguate multiple
capabilities on a single resource.  See section \ref{primarykeys} for details.


\begin{inlinetable}
\small
\begin{tabular}{p{0.28\textwidth}p{0.2\textwidth}p{0.66\textwidth}}
\sptablerule
\multicolumn{3}{l}{\textit{Column names, utypes, ADQL types, and descriptions for the \rtent{rr.capability} table}}\\
\sptablerule
ivoid\hfil\break
\makebox[0pt][l]{\scriptsize\ttfamily xpath:/identifier}&
\footnotesize VARCHAR(*)&
The parent resource.\\
cap\_index\hfil\break
\makebox[0pt][l]{\scriptsize\ttfamily }&
\footnotesize SMALLINT(1)&
An arbitrary identifier of this capability within the resource.\\
cap\_type\hfil\break
\makebox[0pt][l]{\scriptsize\ttfamily xpath:@xsi:type}&
\footnotesize VARCHAR(*)&
The type of capability covered here.\\
cap\_description\hfil\break
\makebox[0pt][l]{\scriptsize\ttfamily xpath:description}&
\footnotesize VARCHAR(*)&
A human-readable description of what this capability provides as part of the over-all service.\\
standard\_id\hfil\break
\makebox[0pt][l]{\scriptsize\ttfamily xpath:@standardID}&
\footnotesize VARCHAR(*)&
A URI for a standard this capability conforms to.\\

\sptablerule
\end{tabular}
\end{inlinetable}


This table should have an explicit primary key made up of
\rtent{ivoid} and \rtent{cap\_index}.
The \rtent{ivoid} column should be
an explicit foreign key into \rtent{resource}.
It is recommended to maintain indexes on at least the 
\rtent{cap\_type} and \rtent{standard\_id} columns.

The following columns MUST be lowercased during ingestion:
\rtent{ivoid}, \rtent{cap\_type}, \rtent{standard\_id}.
Clients are advised to query the \rtent{cap\_description} column
using the \rtent{ivo\_hasword} function.



\subsection{The res\_schema Table}

\label{table_res_schema}

The \rtent{res\_schema} table corresponds to VODataService's
\vorent{schema} element.  It has been renamed to avoid clashes with
the SQL reserved word \texttt{SCHEMA}.

The table has a column \rtent{schema\_index} to disambiguate
multiple schema elements on a single resource.  See section \ref{primarykeys} for details.


\begin{inlinetable}
\small
\begin{tabular}{p{0.28\textwidth}p{0.2\textwidth}p{0.66\textwidth}}
\sptablerule
\multicolumn{3}{l}{\textit{Column names, utypes, ADQL types, and descriptions for the \rtent{rr.res\_schema} table}}\\
\sptablerule
ivoid\hfil\break
\makebox[0pt][l]{\scriptsize\ttfamily xpath:/identifier}&
\footnotesize VARCHAR(*)&
The parent resource.\\
schema\_index\hfil\break
\makebox[0pt][l]{\scriptsize\ttfamily }&
\footnotesize SMALLINT(1)&
An arbitrary identifier for the res\_schema rows belonging to a resource.\\
schema\_description\hfil\break
\makebox[0pt][l]{\scriptsize\ttfamily xpath:description}&
\footnotesize VARCHAR(*)&
A free text description of the tableset explaining in general how all of the tables are related.\\
schema\_name\hfil\break
\makebox[0pt][l]{\scriptsize\ttfamily xpath:name }&
\footnotesize VARCHAR(*)&
A name for the set of tables.\\
schema\_title\hfil\break
\makebox[0pt][l]{\scriptsize\ttfamily xpath:title}&
\footnotesize VARCHAR(*)&
A descriptive, human-interpretable name for the table set.\\
schema\_utype\hfil\break
\makebox[0pt][l]{\scriptsize\ttfamily xpath:utype}&
\footnotesize VARCHAR(*)&
An identifier for a concept in a data model that the data in this schema as a whole represent.\\

\sptablerule
\end{tabular}
\end{inlinetable}


This table should have an explicit primary key made up of
\rtent{ivoid} and \rtent{schema\_index}.  The
\rtent{ivoid}  column should be an explicit foreign key into
\rtent{resource}.

The following columns MUST be lowercased during ingestion:
\rtent{ivoid}, \rtent{schema\_name}, \rtent{schema\_utype}.
Clients are advised to query the \rtent{schema\_description} 
and \rtent{schema\_title} columns
using the the \rtent{ivo\_hasword} function.



\subsection{The res\_table Table}

\label{table_res_table}

The \rtent{res\_table} table models VODataService's
\vorent{table} element.  It has been renamed to avoid name clashes
with the SQL reserved word \texttt{TABLE}.

VODataService 1.0 had a similar element that was a direct child of
resource.  Ingestors should also accept such tables, as there are still
numerous active VODataService 1.0 resources in the Registry at the time
of writing (this is the reason for the alternative in the table xpath).

The table contains a column \rtent{table\_index} to disambiguate
multiple tables on a single resource.  See section \ref{primarykeys} for details.  Note that if the sibling
count is used as implementation of \rtent{table\_index}, the count
must be per resource and \emph{not} per schema, as
\rtent{table\_index} MUST be unique within a resource.


\begin{inlinetable}
\small
\begin{tabular}{p{0.28\textwidth}p{0.2\textwidth}p{0.66\textwidth}}
\sptablerule
\multicolumn{3}{l}{\textit{Column names, utypes, ADQL types, and descriptions for the \rtent{rr.res\_table} table}}\\
\sptablerule
ivoid\hfil\break
\makebox[0pt][l]{\scriptsize\ttfamily xpath:/identifier}&
\footnotesize VARCHAR(*)&
The parent resource.\\
schema\_index\hfil\break
\makebox[0pt][l]{\scriptsize\ttfamily }&
\footnotesize SMALLINT(1)&
Index of the schema this table belongs to, if it belongs to a schema (otherwise NULL).\\
table\_description\hfil\break
\makebox[0pt][l]{\scriptsize\ttfamily xpath:description}&
\footnotesize VARCHAR(*)&
A free-text description of the table's contents.\\
table\_name\hfil\break
\makebox[0pt][l]{\scriptsize\ttfamily xpath:name}&
\footnotesize VARCHAR(*)&
The fully qualified name of the table. This name should include all catalog or schema prefixes needed to distinguish it in a query.\\
table\_index\hfil\break
\makebox[0pt][l]{\scriptsize\ttfamily }&
\footnotesize SMALLINT(1)&
An arbitrary identifier for the tables belonging to a resource.\\
table\_title\hfil\break
\makebox[0pt][l]{\scriptsize\ttfamily xpath:title}&
\footnotesize VARCHAR(*)&
A descriptive, human-interpretable name for the table.\\
table\_type\hfil\break
\makebox[0pt][l]{\scriptsize\ttfamily xpath:@type}&
\footnotesize VARCHAR(*)&
A name for the role this table plays. Recognized values include "output", indicating this table is output from a query; "base\_table", indicating a table whose records represent the main subjects of its schema; and "view", indicating that the table represents a useful combination or subset of other tables. Other values are allowed.\\
table\_utype\hfil\break
\makebox[0pt][l]{\scriptsize\ttfamily xpath:utype}&
\footnotesize VARCHAR(*)&
An identifier for a concept in a data model that the data in this table as a whole represent.\\

\sptablerule
\end{tabular}
\end{inlinetable}


This table should have an explicit primary key made up of
\rtent{ivoid} and \rtent{table\_index}.  The 
\rtent{ivoid} column should be an explicit
foreign key into \rtent{resource}.  It is recommended to
maintain an index on at least the \rtent{table\_description}
column, ideally one suited for queries with \rtent{ivo\_hasword}.

The following columns MUST be lowercased during ingestion:
\rtent{ivoid}, \rtent{table\_name}, \rtent{table\_type},
\rtent{table\_utype}.
Clients are advised to query the \rtent{table\_description}
and \rtent{table\_title}  columns
using the the \rtent{ivo\_hasword} function.



\subsection{The table\_column Table}

\label{table_table_column}

The \rtent{table\_column}  table models the content of VODataService's
\vorent{column} element.  The table has been renamed to avoid
a name clash with the SQL reserved word \texttt{COLUMN}.

Since it is expected that queries for column properties will be
fairly common in advanced queries, it is the column table that has the
unprefixed versions of common member names (name,  ucd,
utype, etc).

The \rtent{flag} column contains a concatenation of all values
of a \vorent{column} element's \vorent{flag} children, separated
by hash characters.  Use the \rtent{ivo\_hashlist\_has} function in
queries against \rtent{flag}.

The \rtent{table\_column} table also includes information from
VODataService's data type concept.  VODataService 1.1 includes several type
systems (VOTable, ADQL, Simple).  The
\rtent{type\_system} column contains the value of the column's 
\vorent{datatype} child, with the VODataService XML prefix fixed
to vs; hence, this column will contain one of NULL,
\texttt{vs:taptype},
\texttt{vs:simpledatatype}, and \texttt{vs:votabletype}.


\begin{inlinetable}
\small
\begin{tabular}{p{0.28\textwidth}p{0.2\textwidth}p{0.66\textwidth}}
\sptablerule
\multicolumn{3}{l}{\textit{Column names, utypes, ADQL types, and descriptions for the \rtent{rr.table\_column} table}}\\
\sptablerule
ivoid\hfil\break
\makebox[0pt][l]{\scriptsize\ttfamily xpath:/identifier}&
\footnotesize VARCHAR(*)&
The parent resource.\\
table\_index\hfil\break
\makebox[0pt][l]{\scriptsize\ttfamily }&
\footnotesize SMALLINT(1)&
Index of the table this column belongs to.\\
name\hfil\break
\makebox[0pt][l]{\scriptsize\ttfamily xpath:name}&
\footnotesize VARCHAR(*)&
The name of the column.\\
ucd\hfil\break
\makebox[0pt][l]{\scriptsize\ttfamily xpath:ucd}&
\footnotesize VARCHAR(*)&
A unified content descriptor that describes the scientific content of the parameter.\\
unit\hfil\break
\makebox[0pt][l]{\scriptsize\ttfamily xpath:unit}&
\footnotesize VARCHAR(*)&
The unit associated with all values in the column.\\
utype\hfil\break
\makebox[0pt][l]{\scriptsize\ttfamily xpath:utype}&
\footnotesize VARCHAR(*)&
An identifier for a role in a data model that the data in this column represents.\\
std\hfil\break
\makebox[0pt][l]{\scriptsize\ttfamily xpath:@std}&
\footnotesize SMALLINT(1)&
If 1, the meaning and use of this parameter is reserved and defined by a standard model. If 0, it represents a database-specific parameter that effectively extends beyond the standard.\\
datatype\hfil\break
\makebox[0pt][l]{\scriptsize\ttfamily xpath:dataType}&
\footnotesize VARCHAR(*)&
The type of the data contained in the column.\\
extended\_schema\hfil\break
\makebox[0pt][l]{\scriptsize\ttfamily xpath:dataType/@extendedSchema}&
\footnotesize VARCHAR(*)&
An identifier for the schema that the value given by the extended attribute is drawn from.\\
extended\_type\hfil\break
\makebox[0pt][l]{\scriptsize\ttfamily xpath:dataType/@extendedType}&
\footnotesize VARCHAR(*)&
A custom type for the values this column contains.\\
arraysize\hfil\break
\makebox[0pt][l]{\scriptsize\ttfamily xpath:dataType/@arraysize}&
\footnotesize VARCHAR(*)&
The shape of the array that constitutes the value, e.g., 4, *, 4*, 5x4, or 5x*, as specified by VOTable.\\
delim\hfil\break
\makebox[0pt][l]{\scriptsize\ttfamily xpath:dataType/@delim}&
\footnotesize VARCHAR(*)&
The string that is used to delimit elements of an array value when arraysize is not '1'.\\
type\_system\hfil\break
\makebox[0pt][l]{\scriptsize\ttfamily xpath:dataType/@xsi:type}&
\footnotesize VARCHAR(*)&
The type system used, as a QName with a canonical prefix; this will ususally be one of vs:simpledatatype, vs:votabletype, and vs:taptype.\\
flag\hfil\break
\makebox[0pt][l]{\scriptsize\ttfamily xpath:flag}&
\footnotesize VARCHAR(*)&
Hash-separated keywords representing traits of the column. Recognized values include "indexed", "primary", and "nullable".\\
column\_description\hfil\break
\makebox[0pt][l]{\scriptsize\ttfamily xpath:description}&
\footnotesize VARCHAR(*)&
A free-text description of the column's contents.\\

\sptablerule
\end{tabular}
\end{inlinetable}


The pair \rtent{ivoid}, \rtent{table\_index} should be an
explicit foreign key into \rtent{res\_table}.  It is recommended to
maintain indexes on at least the \rtent{column\_description},
\rtent{name}, \rtent{ucd}, and \rtent{utype} columns,
where the index on \rtent{column\_description} should ideally be able
to handle queries using \rtent{ivo\_hasword}.

The following columns MUST be lowercased during ingestion:
\rtent{ivoid}, \rtent{name}, \rtent{ucd},
\rtent{utype}, \rtent{datatype}, \rtent{type\_system}.
The boolean value of the column's \rtent{std} attribute must be
converted to 0 (False), 1 (True), or NULL (not given) on ingestion.
Clients are advised to query the \rtent{description}
column using the \rtent{ivo\_hasword} function, and to query
the \rtent{flag} column using the \rtent{ivo\_hashlist\_has}
function.



\subsection{The interface Table}

\label{table_interface}

The \rtent{interface} table subsumes both the
\vorent{vr:Interface} and \vorent{vr:access\-URL} types from
VOResource.  The integration of \vorent{access\-URL}  into
the \rtent{interface}  table means that an interface in the
relational registry can only have one access URL, where in VOResource it
can have many.  In practice, this particular VOResource capability has
not been used by registry record authors.  Since access URLs are
probably the item most queried for, it seems warranted to save one
indirection when querying for them.

This specification deprecates the \texttt{maxOccurs=\dquote unbounded"}
in the definition of \vorent{Interface}'s \vorent{accessURL}
child in the XML schema
\texttt{http://www.ivoa.net/xml/VOResource/v1.0}; in future versions
of VOResource, implementations can expect this to be
\texttt{maxOccurs="1"}.  Meanwhile, implementation behavior in the
presence of multiple access URLs in an interface is undefined.

The table contains a column \rtent{intf\_index} to disambiguate
multiple interfaces of one resource. See section \ref{primarykeys} for details.

In VOResource, interfaces can have zero or more \vorent{securityMethod}
children to convey support for authentication and authorization methods.  At
the time of writing, only a \vorent{standardId} attribute is defined on
these, and the Registry only contains a single resource record using
\vorent{securityMethod}.  It was therefore decided to map this information
into the \rtent{res\_detail} table using the detail xpath
\texttt{/capability/interface/securityMethod/@standardID}, with
\rtent{cap\_index} set so the embedding capability is referenced;
this implies that RegTAP does not allow distinguishing between
interfaces for security methods.  If \vorent{securityMethod} will
come into common use as the VO evolves, this design will be reconsidered
if reliable RegTAP-based discovery of access URLs in multi-interface
multi-authentication scenarios is found necessary.

The \rtent{query\_type} column is a hash-joined list (analogous
to \rtent{waveband}, etc. in the resource table), as
the XML schema allows listing up to two request methods.

This table only contains interface elements from within capabilities.
Interface elements in StandardsRegExt records are ignored in the
relational registry,
and they must not be inserted in this table, since doing so would disturb
the foreign key from interface into capability.  In other words,
the relational registry requires every interface to have a parent capability.

Analogous to \rtent{resource.res\_type}, the
\rtent{intf\_type} column contains type names; VOResource extensions
can define new types here, but at the time of writing, the following
types are mentioned in IVOA-recommended schemata:

\begin{description}
\item[vs:paramhttp]A service invoked via an HTTP query, usually with some form of
structured parameters. This type is used for interfaces speaking
standard IVOA protocols.
\item[vr:webbrowser]A (form-based) interface intended to be accessed interactively by a
user via a web browser.
\item[vg:oaihttp]A standard OAI PMH interface using HTTP queries with form-urlencoded
parameters.
\item[vg:oaisoap]A standard OAI PMH interface using a SOAP Web Service
interface.
\item[vr:webservice]A Web Service that is describable by a WSDL document.

\end{description}


\begin{inlinetable}
\small
\begin{tabular}{p{0.28\textwidth}p{0.2\textwidth}p{0.66\textwidth}}
\sptablerule
\multicolumn{3}{l}{\textit{Column names, utypes, ADQL types, and descriptions for the \rtent{rr.interface} table}}\\
\sptablerule
ivoid\hfil\break
\makebox[0pt][l]{\scriptsize\ttfamily xpath:/identifier}&
\footnotesize VARCHAR(*)&
The parent resource.\\
cap\_index\hfil\break
\makebox[0pt][l]{\scriptsize\ttfamily }&
\footnotesize SMALLINT(1)&
The index of the parent capability.\\
intf\_index\hfil\break
\makebox[0pt][l]{\scriptsize\ttfamily }&
\footnotesize SMALLINT(1)&
An arbitrary identifier for the interfaces of a resource.\\
intf\_type\hfil\break
\makebox[0pt][l]{\scriptsize\ttfamily xpath:@xsi:type}&
\footnotesize VARCHAR(*)&
The type of the interface (vr:webbrowser, vs:paramhttp, etc).\\
intf\_role\hfil\break
\makebox[0pt][l]{\scriptsize\ttfamily xpath:@role}&
\footnotesize VARCHAR(*)&
An identifier for the role the interface plays in the particular capability. If the value is equal to "std" or begins with "std:", then the interface refers to a standard interface defined by the standard referred to by the capability's standardID attribute.\\
std\_version\hfil\break
\makebox[0pt][l]{\scriptsize\ttfamily xpath:@version}&
\footnotesize VARCHAR(*)&
The version of a standard interface specification that this interface complies with. When the interface is provided in the context of a Capability element, then the standard being refered to is the one identified by the Capability's standardID element.\\
query\_type\hfil\break
\makebox[0pt][l]{\scriptsize\ttfamily xpath:queryType}&
\footnotesize VARCHAR(*)&
Hash-joined list of expected HTTP method (get or post) supported by the service.\\
result\_type\hfil\break
\makebox[0pt][l]{\scriptsize\ttfamily xpath:resultType}&
\footnotesize VARCHAR(*)&
The MIME type of a document returned in the HTTP response.\\
wsdl\_url\hfil\break
\makebox[0pt][l]{\scriptsize\ttfamily xpath:wsdlURL}&
\footnotesize VARCHAR(*)&
The location of the WSDL that describes this Web Service. If NULL, the location can be assumed to be the accessURL with '?wsdl' appended.\\
url\_use\hfil\break
\makebox[0pt][l]{\scriptsize\ttfamily xpath:accessURL/@use}&
\footnotesize VARCHAR(*)&
A flag indicating whether this should be interpreted as a base URL ('base'), a full URL ('full'), or a URL to a directory that will produce a listing of files ('dir').\\
access\_url\hfil\break
\makebox[0pt][l]{\scriptsize\ttfamily xpath:accessURL}&
\footnotesize VARCHAR(*)&
The URL at which the interface is found.\\

\sptablerule
\end{tabular}
\end{inlinetable}


 This table should have the pair \rtent{ivoid}, \rtent{cap\_index}
as an explicit foreign key into
\rtent{capability}, and the pair \rtent{ivoid}, and
\rtent{intf\_index} as an explicit primary key. Additionally, it
is recommended to maintain an index on at least the
\rtent{intf\_type} column.

The following columns MUST be lowercased during ingestion:
\rtent{ivoid}, \rtent{intf\_type}, \rtent{intf\_role},
\rtent{std\_version}, \rtent{query\_type},
\rtent{result\_type}, \rtent{url\_use}.
Clients are advised to query \rtent{query\_type} using the the
\rtent{ivo\_hashlist\_has} function.



\subsection{The intf\_param Table}

\label{table_intf_param}

The \rtent{intf\_param} table keeps information on the parameters
available on interfaces.  It is therefore closely related to
\rtent{table\_column}, but the differences between the two are
significant enough to warrant a separation between the two tables.
Since the names of common column attributes are used where applicable in
both tables (e.g., name, ucd, etc), the two tables cannot be (naturally)
joined.


\begin{inlinetable}
\small
\begin{tabular}{p{0.28\textwidth}p{0.2\textwidth}p{0.66\textwidth}}
\sptablerule
\multicolumn{3}{l}{\textit{Column names, utypes, ADQL types, and descriptions for the \rtent{rr.intf\_param} table}}\\
\sptablerule
ivoid\hfil\break
\makebox[0pt][l]{\scriptsize\ttfamily xpath:/identifier}&
\footnotesize VARCHAR(*)&
The parent resource.\\
intf\_index\hfil\break
\makebox[0pt][l]{\scriptsize\ttfamily }&
\footnotesize SMALLINT(1)&
The index of the interface this parameter belongs to.\\
name\hfil\break
\makebox[0pt][l]{\scriptsize\ttfamily xpath:name}&
\footnotesize VARCHAR(*)&
The name of the parameter.\\
ucd\hfil\break
\makebox[0pt][l]{\scriptsize\ttfamily xpath:ucd}&
\footnotesize VARCHAR(*)&
A unified content descriptor that describes the scientific content of the parameter.\\
unit\hfil\break
\makebox[0pt][l]{\scriptsize\ttfamily xpath:unit}&
\footnotesize VARCHAR(*)&
The unit associated with all values in the parameter.\\
utype\hfil\break
\makebox[0pt][l]{\scriptsize\ttfamily xpath:utype}&
\footnotesize VARCHAR(*)&
An identifier for a role in a data model that the data in this parameter represents.\\
std\hfil\break
\makebox[0pt][l]{\scriptsize\ttfamily xpath:@std}&
\footnotesize SMALLINT(1)&
If 1, the meaning and use of this parameter is reserved and defined by a standard model. If 0, it represents a database-specific parameter that effectively extends beyond the standard.\\
datatype\hfil\break
\makebox[0pt][l]{\scriptsize\ttfamily xpath:dataType}&
\footnotesize VARCHAR(*)&
The type of the data contained in the parameter.\\
extended\_schema\hfil\break
\makebox[0pt][l]{\scriptsize\ttfamily xpath:dataType/@extendedSchema}&
\footnotesize VARCHAR(*)&
An identifier for the schema that the value given by the extended attribute is drawn from.\\
extended\_type\hfil\break
\makebox[0pt][l]{\scriptsize\ttfamily xpath:dataType/@extendedType}&
\footnotesize VARCHAR(*)&
A custom type for the values this parameter contains.\\
arraysize\hfil\break
\makebox[0pt][l]{\scriptsize\ttfamily xpath:dataType/@arraysize}&
\footnotesize VARCHAR(*)&
The shape of the array that constitutes the value, e.g., 4, *, 4*, 5x4, or 5x*, as specified by VOTable.\\
delim\hfil\break
\makebox[0pt][l]{\scriptsize\ttfamily xpath:dataType/@delim}&
\footnotesize VARCHAR(*)&
The string that is used to delimit elements of an array value when arraysize is not '1'.\\
param\_use\hfil\break
\makebox[0pt][l]{\scriptsize\ttfamily xpath:@use}&
\footnotesize VARCHAR(*)&
An indication of whether this parameter is required to be provided for the application or service to work properly (one of required, optional, ignored, or NULL).\\
param\_description\hfil\break
\makebox[0pt][l]{\scriptsize\ttfamily xpath:description}&
\footnotesize VARCHAR(*)&
A free-text description of the parameter's contents.\\

\sptablerule
\end{tabular}
\end{inlinetable}


The pair \rtent{ivoid}, \rtent{intf\_index} should be an explicit
foreign key into \rtent{interface}.

The remaining requirements and conventions are as per
section \ref{table_table_column}
where applicable, and \rtent{param\_description} taking the role
of \rtent{column\_description}.



\subsection{The relationship Table}

\label{table_relationship}

The relationship element is a slight denormalization of the
\vorent{vr:Relation\-ship} type: Whereas in VOResource, a single
relationship element can take several IVORNs, in the relational model,
the pairs are stored directly.  It is straightforward to translate
between the two representations in the database ingestor.


\begin{inlinetable}
\small
\begin{tabular}{p{0.28\textwidth}p{0.2\textwidth}p{0.66\textwidth}}
\sptablerule
\multicolumn{3}{l}{\textit{Column names, utypes, ADQL types, and descriptions for the \rtent{rr.relationship} table}}\\
\sptablerule
ivoid\hfil\break
\makebox[0pt][l]{\scriptsize\ttfamily xpath:/identifier}&
\footnotesize VARCHAR(*)&
The parent resource.\\
relationship\_type\hfil\break
\makebox[0pt][l]{\scriptsize\ttfamily xpath:relationshipType}&
\footnotesize VARCHAR(*)&
The named type of relationship; this can be mirror-of, service-for, served-by, derived-from, related-to.\\
related\_id\hfil\break
\makebox[0pt][l]{\scriptsize\ttfamily xpath:relatedResource/@ivo-id}&
\footnotesize VARCHAR(*)&
The IVOA identifier for the resource referred to.\\
related\_name\hfil\break
\makebox[0pt][l]{\scriptsize\ttfamily xpath:relatedResource}&
\footnotesize VARCHAR(*)&
The name of resource that this resource is related to.\\

\sptablerule
\end{tabular}
\end{inlinetable}


The \rtent{ivoid} column should be an explicit foreign key into the
\rtent{resource} table.  You should index at least the
\rtent{related\_id} column.

The following columns MUST be lowercased during ingestion:
\rtent{ivoid}, \rtent{relationship\_type},
\rtent{related\_id}.



\subsection{The validation Table}

\label{table_validation}

The \rtent{validation} table subsumes the
\vorent{vr:validationLevel}-typed members of both \vorent{vr:Resource}
and \vorent{vr:Capability}.

If the \rtent{cap\_index} column is \texttt{NULL}, the
validation comprises the entire resource.  Otherwise, only the
referenced capability has been validated.

While it is recommended that harvesters only accept resource records
from their originating registries, it is valuable to gather validation
results from various sources.  Hence, harvesters for the relational
registry may choose to obtain validation data from the OAI-PMH endpoints
of various registries by not harvesting just for the ivo\_managed set and
generate \rtent{rr.validation} rows from these records.  This can
trigger potentially problematic behavior when the original registry
updates  its resource record in that naive implementations will lose all
third-party validation rows; this may actually be the correct behavior,
since an update of the registry record might very well indicate
validation-relevant changes in the underlying services.  Implementations
are free to handle or ignore validation results as they see fit, and
they may add validation results of their own.

The validation levels are defined in \citet{std:RM} and
currently range from 0 (description stored in a registry) to 
4 (inspected by a human to be technically and scientifically
correct).


\begin{inlinetable}
\small
\begin{tabular}{p{0.28\textwidth}p{0.2\textwidth}p{0.66\textwidth}}
\sptablerule
\multicolumn{3}{l}{\textit{Column names, utypes, ADQL types, and descriptions for the \rtent{rr.validation} table}}\\
\sptablerule
ivoid\hfil\break
\makebox[0pt][l]{\scriptsize\ttfamily xpath:/identifier}&
\footnotesize VARCHAR(*)&
The parent resource.\\
validated\_by\hfil\break
\makebox[0pt][l]{\scriptsize\ttfamily xpath:validationLevel/@validatedBy}&
\footnotesize VARCHAR(*)&
The IVOA ID of the registry or organisation that assigned the validation level.\\
val\_level\hfil\break
\makebox[0pt][l]{\scriptsize\ttfamily xpath:validationLevel}&
\footnotesize SMALLINT(1)&
A numeric grade describing the quality of the resource description, when applicable, to be used to indicate the confidence an end-user can put in the resource as part of a VO application or research study.\\
cap\_index\hfil\break
\makebox[0pt][l]{\scriptsize\ttfamily }&
\footnotesize SMALLINT(1)&
If non-NULL, the validation only refers to the capability referenced here.\\

\sptablerule
\end{tabular}
\end{inlinetable}


The \rtent{ivoid} column should be an explicit foreign key into
\rtent{resource}.  Note, however, that \rtent{ivoid},
\rtent{cap\_index} is \emph{not} a foreign key into \rtent{capability}
since \rtent{cap\_index} may be \texttt{NULL} (in case the validation
addresses the entire resource).

The following columns MUST be lowercased during ingestion:
\rtent{ivoid}, \rtent{validated\_by}.



\subsection{The res\_date Table}

\label{table_res_date}

The \rtent{res\_date} table contains information gathered from
\vorent{vr:Curation}'s date children.


\begin{inlinetable}
\small
\begin{tabular}{p{0.28\textwidth}p{0.2\textwidth}p{0.66\textwidth}}
\sptablerule
\multicolumn{3}{l}{\textit{Column names, utypes, ADQL types, and descriptions for the \rtent{rr.res\_date} table}}\\
\sptablerule
ivoid\hfil\break
\makebox[0pt][l]{\scriptsize\ttfamily xpath:/identifier}&
\footnotesize VARCHAR(*)&
The parent resource.\\
date\_value\hfil\break
\makebox[0pt][l]{\scriptsize\ttfamily xpath:date}&
\footnotesize TIMESTAMP(1)&
A date associated with an event in the life cycle of the resource.\\
value\_role\hfil\break
\makebox[0pt][l]{\scriptsize\ttfamily xpath:date/@role}&
\footnotesize VARCHAR(*)&
A string indicating what the date refers to, e.g., created, availability, updated.\\

\sptablerule
\end{tabular}
\end{inlinetable}


The \rtent{ivoid} column should be an explicit foreign key into
\rtent{resource}.

The following columns MUST be lowercased during ingestion:
\rtent{ivoid}, \rtent{value\_role}.



\subsection{The res\_detail Table}

\label{table_res_detail}

The \rtent{res\_detail} table is the relational registry's primary means for
extensibility as well as a fallback for less-used simple
metadata.  Conceptually, it stores triples of resource entity
references, resource xpaths,
and values, where resource entities can be resource records themselves
or capabilities.  Thus, metadata with values that are either
string-valued or sets of strings can be represented in this table.

As long as the metadata that needs to be represented in the
relational registry for new VOResource extensions is simple enough, no changes to the schema defined
here will be necessary as these are introduced.  Instead, the extension itself simply defines
new xpaths to be added in \rtent{res\_detail}.

Some complex metadata -- \vorent{tr:languageFeature} or
\vorent{vstd:key} being examples -- cannot be kept in this table.
If a representation of such information in the relational registry is
required, this standard will need to be changed.

Appendix \ref{d_u_list} gives a list
of resource xpaths from the registry extensions
that were recommendations at the time of writing.  
For the resource xpaths marked with an exclamation mark there,
xpath/value pairs MUST be generated whenever the corresponding
metadata items are given in a resource record.
For the remaining resource xpaths, these pairs should be provided if
convenient; they mostly concern test queries and other curation-type
information that, while unlikely to be useful to normal users, may be
relevant to curation-type clients that, e.g., ascertain the continued 
operation of services.

Some detail values must be interpreted case-insensitively; this
concerns, in particular, IVORNs like the TAP data model type.  For other
rows -- the test queries are immediate examples -- , changing the case
will likely break the data.  In order to avoid having to give and
implement case normalization rules by detail xpath, no case normalization 
is done on detail values at all, and users and clients will have to use
the \rtent{ivo\_nocasematch} user defined function (see section
\ref{adqludf}) when locating
case-insensitive values.  For the resource xpaths given in Appendix \ref{d_u_list}, this concerns all items with xpaths ending
in \texttt{@ivo-id}.

Individual
ingestors
MAY choose to expose additional metadata using other utypes, provided
they are formed according to the rules in 
section \ref{vorutypes} (this rule is intended
to minimize the risk of later clashes).

In addition to the metadata listed in this specification, 
metadata defined in future
IVOA-approved VOResource extensions MUST or SHOULD be present in
\rtent{res\_details} as the extensions require it.


\begin{inlinetable}
\small
\begin{tabular}{p{0.28\textwidth}p{0.2\textwidth}p{0.66\textwidth}}
\sptablerule
\multicolumn{3}{l}{\textit{Column names, utypes, ADQL types, and descriptions for the \rtent{rr.res\_detail} table}}\\
\sptablerule
ivoid\hfil\break
\makebox[0pt][l]{\scriptsize\ttfamily xpath:/identifier}&
\footnotesize VARCHAR(*)&
The parent resource.\\
cap\_index\hfil\break
\makebox[0pt][l]{\scriptsize\ttfamily }&
\footnotesize SMALLINT(1)&
The index of the parent capability; if NULL the xpath-value pair describes a member of the entire resource.\\
detail\_xpath\hfil\break
\makebox[0pt][l]{\scriptsize\ttfamily }&
\footnotesize VARCHAR(*)&
The xpath of the data item.\\
detail\_value\hfil\break
\makebox[0pt][l]{\scriptsize\ttfamily }&
\footnotesize VARCHAR(*)&
(Atomic) value of the member.\\

\sptablerule
\end{tabular}
\end{inlinetable}


The \rtent{ivoid} column should be an explicit foreign key into
\rtent{resource}.  It is recommended to maintain indexes on
at least the columns
\rtent{detail\_xpath} and \rtent{detail\_value}, where the
index on \rtent{detail\_value} should ideally work for both direct
comparisons and searches using the \rtent{ivo\_nocasematch}
function.

The following column MUST be lowercased during ingestion:
\rtent{ivoid}.  Clients are advised to
use the \rtent{ivo\_nocasematch} function to search in
\rtent{detail\_value} if the values are to be compared
case-insensitively (e.g., all IVORNs).




\section{ADQL User Defined Functions}

\label{adqludf}

TAP Servers implementing the 
\texttt{ivo://ivoa.net/std/RegTAP\#1.0} data model MUST 
implement the following four functions in their ADQL 2.0 language,
with signatures written as recommended in \citet{std:TAPREGEXT}:

\begin{bigdescription}
\term[\small\texttt{ivo\_nocasematch(value VARCHAR(*), pat VARCHAR(*))->INTEGER}]
The function returns 1 if \texttt{pat}  matches
\texttt{value}, 0 otherwise.
\texttt{pat}  is defined as for the SQL LIKE operator, but the
match is performed case-insensitively.

\term[\small\texttt{ivo\_hasword(haystack VARCHAR(*), needle VARCHAR(*)) -> INTEGER}]The function takes two strings and returns 1 if the second is
contained in the first one in a ``word'' sense, i.e., delimited by
non-letter characters or the beginning or end of the string, where case
is ignored.
Additionally, servers MAY employ techniques to improve recall, in
particular stemming.  Registry clients must hence expect different results
from different servers when using \rtent{ivo\_hasword}; for such
queries trying them on multiple registries may improve recall.
\term[\small\texttt{ivo\_hashlist\_has(hashlist VARCHAR(*), item VARCHAR(*)) -> INTEGER}]The function takes two strings; the first is a list of words not
containing the hash sign (\#), concatenated by hash signs, the second is
a word not containing the hash sign.  It returns 1 if, compared
case-insensitively, the second argument is in the list of words encoded in
the first argument.  The behavior for second
arguments containing a hash sign is undefined.
\term[\small\texttt{ivo\_string\_agg(expr VARCHAR(*), deli VARCHAR(*)) -> VARCHAR(*)}]An aggregate function returning all values of
\texttt{expr} within a GROUP concatenated with 
\texttt{delim}.  NULLs in the aggregate do not contribute, an empty 
aggregate yields an empty string.

\end{bigdescription}

Reference implementations of the four functions for the PostgreSQL
database system are given in Appendix \ref{appPGDefs}.



\section{Common Queries to the Relational Registry}

\label{sample_queries}

This section contains sample queries to the relational registry,
mostly contributed as use cases by various members of the IVOA Registry
working group.  They are intended as an aid in designing relational
registry queries, in particular for users new to the data model.

When locating access URLs for capabilities of standard services, these
sample queries look for interfaces of type \vorent{vs:ParamHTTP} within
the embedding capabilities.  This is not strictly as intended by
VOResource, which has the special \vorent{role} attribute to mark the
interface on which a standard protocol is exposed within a capability.
In RegTAP, this method of locating standard interfaces would be effected
by constraining \texttt{intf\_role LIKE 'std\%'}.  In actual VO
practice, too many standard interfaces are lacking a proper declaration
of their role, and actual clients locate the standard interfaces as
given here.  Following widespread practice client designers are
encouraged to compare against the interface types rather than rely on
\vorent{interface/@role}, and resource record authors should make sure
clients can discover standard interfaces both by the interfaces' roles
and types.

\subsection{TAP accessURLs}
\textbf{Problem:} Find all TAP services; return their accessURLs

As the capability type is in 
\rtent{rr.capability}, whereas the access URL can be
found from 
\rtent{rr.interface}, this requires
a join:

\begin{lstlisting}[language=SQL,flexiblecolumns=true]
SELECT ivoid, access_url 
FROM rr.capability
NATURAL JOIN rr.interface
WHERE standard_id like 'ivo://ivoa.net/std/tap%'
  AND intf_type='vs:paramhttp'
\end{lstlisting}

Other \rtent{standard\_id}s relevant here include:

\begin{itemize}

\item \texttt{ivo://ivoa.net/std/registry} for OAI-PMH services,{}

\item \texttt{ivo://ivoa.net/std/sia} for SIA services,{}

\item \texttt{ivo://ivoa.net/std/conesearch} for SCS services,
and{}

\item \texttt{ivo://ivoa.net/std/ssa} for SSA services.{}

\end{itemize}

\subsection{Image Services with Spirals}

\textbf{Problem:} Find all SIA services that might have spiral
galaxies

This is somewhat tricky since it is probably hard to image a part
of the sky guaranteed not to have some, possibly distant, spiral galaxy
in it.  However, translating the intention into ``find all SIA services
that mention spiral in either the subject (from 
\rtent{rr.res\_subject}), the description, or the
title (which are in 
\rtent{rr.resource})'', 
the query would become:

\begin{lstlisting}[language=SQL,flexiblecolumns=true]
SELECT ivoid, access_url 
FROM rr.capability 
  NATURAL JOIN rr.resource
  NATURAL JOIN rr.interface
  NATURAL JOIN rr.res_subject
WHERE standard_id='ivo://ivoa.net/std/sia'
  AND intf_type='vs:paramhttp'
  AND (
    1=ivo_nocasematch(res_subject, '%spiral%')
    OR 1=ivo_hasword(res_description, 'spiral')
    OR 1=ivo_hasword(res_title, 'spiral'))
\end{lstlisting}

\subsection{Infrared Image Services}

\textbf{Problem:} Find all SIA services that provide infrared
images

The waveband information in 
\rtent{rr.resource} 
comes in hash-separated atoms (which can be
radio, millimeter, infrared, optical, uv, euv, x-ray, or gamma-ray).
For matching those, use the \rtent{ivo\_hashlist\_has} function as
below.  The access URL and the service standard come from 
\rtent{rr.interface} and 
\rtent{rr.capability}, respectively.

\begin{lstlisting}[language=SQL,flexiblecolumns=true]
SELECT ivoid, access_url 
FROM rr.capability 
  NATURAL JOIN rr.resource
  NATURAL JOIN rr.interface
WHERE standard_id='ivo://ivoa.net/std/sia'
  AND intf_type='vs:paramhttp'
  AND 1=ivo_hashlist_has('infrared', waveband)
\end{lstlisting}

\subsection{Catalogs with Redshifts}
\textbf{Problem:} Find all searchable catalogs (i.e., cone search
services) that provide a column containing redshifts

Metadata on columns exposed by a service are contained in 
\rtent{rr.table\_column}.  Again, this table can be
naturally joined with 
\rtent{rr.capability} and 
\rtent{rr.interface}.
:
\begin{lstlisting}[language=SQL,flexiblecolumns=true]
SELECT ivoid, access_url 
FROM rr.capability 
  NATURAL JOIN rr.table_column
  NATURAL JOIN rr.interface 
WHERE standard_id='ivo://ivoa.net/std/conesearch'
  AND intf_type='vs:paramhttp'
  AND ucd='src.redshift'
\end{lstlisting}

Sometimes you want to match a whole set of ucds.  Frequently the
simple regular expressions of SQL will help, as in 
\texttt{AND ucd LIKE 'pos.parallax\%'}.  When that is not enough, 
use boolean OR expressions 

\subsection{Names from an Authority}

\textbf{Problem:} Find all the resources published by a certain
authority

An ``authority'' within the VO is something that hands out identifiers.
You can tell what authority a record came from by looking at the ``host
part'' of the IVO identifier, most naturally obtained from 
\rtent{rr.resource}.  Since ADQL cannot actually parse
URIs, we make do with simple string matching:

\begin{lstlisting}[language=SQL,flexiblecolumns=true]
SELECT ivoid 
FROM rr.resource
WHERE ivoid LIKE 'ivo://org.gavo.dc%'
\end{lstlisting}

\subsection{Records Published by X}

\textbf{Problem:} What registry records are there from a given
publisher?

This uses the 
\rtent{rr.res\_role}
 table both to
match names (in this case, a publisher that has ``gavo'' in its name) and
to ascertain the named entity actually publishes the resource (rather
than, e.g., just being the contact in case of trouble).  The result is a
list of ivoids in this case.  You could join this with any other
table in the relational registry to find out more about these
services.

\begin{lstlisting}[language=SQL,flexiblecolumns=true]
SELECT ivoid 
FROM rr.res_role
WHERE 1=ivo_nocasematch(role_name, '%gavo%')
  AND base_role='publisher'
\end{lstlisting}

or, if the publisher actually gives its ivo-id in the registry
records,

\begin{lstlisting}[language=SQL,flexiblecolumns=true]
SELECT ivoid 
FROM rr.res_role
WHERE role_ivoid='ivo://ned.ipac/ned'
  AND base_role='publisher'
\end{lstlisting}

\subsection{Records from Registry}

\textbf{Problem:} What registry records are
there originating from registry X?

This is mainly a query interesting for registry maintainers.  Still,
it is a nice example for joining with the
\rtent{rr.res\_detail} table, in this case to
first get a list of all authorities managed by the CDS registry.

\begin{lstlisting}[language=SQL,flexiblecolumns=true]
SELECT ivoid FROM rr.resource
RIGHT OUTER JOIN (
  SELECT 'ivo://' || detail_value || '%' AS pat
  FROM rr.res_detail
  WHERE detail_xpath='/managedAuthority' 
    AND ivoid='ivo://cds.vizier/registry') 
  AS authpatterns
ON (1=ivo_nocasematch(resource.ivoid, authpatterns.pat))
\end{lstlisting}

\subsection{Locate RegTAP services}

\textbf{Problem:} Find all TAP endpoints offering the relational
registry

This is the discovery query for RegTAP services themselves;  note how
this combines 
\rtent{rr.res\_detail} pairs with
\rtent{rr.capability}
and 
\rtent{rr.interface} to locate the desired protocol
endpoints.  As clients should not usally be concerned with minor
versions of protocols unless  they rely on additions made in later
versions, this query will return endpoints supporting ``version 1'' rather
than exactly version 1.0.

\begin{lstlisting}[language=SQL,flexiblecolumns=true]
SELECT access_url
FROM rr.interface
NATURAL JOIN rr.capability
NATURAL JOIN rr.res_detail
WHERE standard_id='ivo://ivoa.net/std/tap'
  AND intf_type='vs:paramhttp'
  AND detail_xpath='/capability/dataModel/@ivo-id'
  AND 1=ivo_nocasematch(detail_value, 
    'ivo://ivoa.net/std/regtap#1.%')
\end{lstlisting}

\subsection{TAP with Physics}

\textbf{Problem:} Find all TAP services
exposing a table with certain features

``Certain features'' could be ``have some word in their description
and having a column with a certain UCD''.  Either way, this kind of query
fairly invariably combines the usual 
\rtent{rr.capability} and 
\rtent{rr.interface}
 for service location with
\rtent{rr.table\_column}
 for the column metadata
and 
\rtent{rr.res\_table} for properties of tables.

\begin{lstlisting}[language=SQL,flexiblecolumns=true]
SELECT ivoid, access_url, name, ucd, column_description
FROM rr.capability 
  NATURAL JOIN rr.interface
  NATURAL JOIN rr.table_column
  NATURAL JOIN rr.res_table
WHERE standard_id='ivo://ivoa.net/std/tap'
  AND intf_type='vs:paramhttp'
  AND 1=ivo_hasword(table_description, 'quasar')
  AND ucd='phot.mag;em.opt.v'
\end{lstlisting}

\subsection{Theoretical SSA}

\textbf{Problem:} Find all SSAP services that
provide theoretical spectra.

The metadata required to solve this problem is found in the SSAP
registry extension and is thus kept in 
\rtent{rr.res\_detail}:

\begin{lstlisting}[language=SQL,flexiblecolumns=true]
SELECT access_url 
FROM rr.res_detail 
  NATURAL JOIN rr.capability 
  NATURAL JOIN rr.interface 
WHERE detail_xpath='/capability/dataSource' 
  AND intf_type='vs:paramhttp'
  AND standard_id='ivo://ivoa.net/std/ssa'
  AND detail_value='theory'
\end{lstlisting}

\subsection{Find Contact Persons}

\textbf{Problem:} The service at
\texttt{http://dc.zah.uni-heidelberg.de/tap} is down, who can
fix it?

This uses the \rtent{rr.res\_role} table and returns all information on
it based on the IVORN of a service that in turn was obtained from
\rtent{rr.interface}.  You could restrict to the actual technical
contact person by requiring \texttt{base\_role='contact'}.

\begin{lstlisting}[language=SQL,flexiblecolumns=true]
SELECT DISTINCT base_role, role_name, email 
FROM rr.res_role 
  NATURAL JOIN rr.interface 
WHERE access_url='http://dc.zah.uni-heidelberg.de/tap'
\end{lstlisting}

\subsection{Related Capabilities}

\textbf{Problem:} Get the capabilities of all services serving a
specific resource (typically, a data collection).

In the VO, data providers can register pure data collections without
access options (or just furnished with a link to a download).  They can
then declare that their data can be ``served-by'' some other resource,
typically a TAP service or some collective service for a number of
instruments.  To locate the access options to the data itself, inspect
\rtent{rr.relationship} and use it to select records
from 
\rtent{rr.capability}.

\begin{lstlisting}[language=SQL,flexiblecolumns=true]
SELECT * 
FROM rr.relationship AS a
  JOIN rr.capability AS b 
    ON (a.related_id=b.ivoid) 
WHERE relationship_type='served-by'
\end{lstlisting}



\appendix

\section{XPaths for res\_details}

\label{d_u_list}

This appendix defines the \rtent{res\_details}
table (see section \ref{table_res_detail} for
details) by giving
xpaths for which xpath/value pairs MUST (where marked with an
exclamation mark) or SHOULD be given if the
corresponding data is present in the resource records.  This list is
normative for metadata defined in IVOA recommendations current as of the
publication of this document (see section \ref{rolewithinivoa}).  
As laid down in section \ref{table_res_detail}, 
new VOResource extensions or new
versions of existing VOResource extensions may amend this list.

In case there are conflicts between this list and xpaths derived 
from schema files using the rules given in section \ref{vorutypes}, the conflict must be considered due to an
editorial oversight in the preparation of this list, and the xpaths from the
schema files are normative.  Errata to this list will be issued in such
cases.

The xpaths are sufficient for locating the respective metadata as per
section \ref{vorutypes}.  With the explanations we
give the canonical prefixes for the XML namespaces from which the items
originate, which is where further information can be found.

\begin{description}
\item[/accessURL (!)]For data collection resources, this is
the URL that can be used to download the data contained.  Do
\emph{not} enter accessURLs from interfaces into res\_detail (vs).
\item[/capability/executionDuration/hard]The hard run time limit, given in seconds (tr).
\item[/capability/complianceLevel]The category indicating the level to which this instance complies with the SSA standard (ssap).
\item[/capability/creationType (!)]The category that describes the process used to produce the dataset; one of archival, cutout, filtered, mosaic, projection, specialExtraction, catalogExtraction (ssap).
\item[/capability/dataModel (!)]The short, human-readable name of a data model supported by a TAP service; for most applications, clients should rather constrain /capability/dataModel/@ivo-id (tr).  
\item[/capability/dataModel/@ivo-id (!)]The IVORN of the data model supported by a TAP service (tr).
\item[/capability/dataSource (!)]The category specifying where the data originally came from; one of survey, pointed, custom, theory, artificial (ssap).
\item[/capability/defaultMaxRecords (!)]The largest number of records that the service will return when the MAXREC parameter is not specified in the query input (ssap).
\item[/capability/executionDuration/default]The run time limit for newly-created jobs, given in seconds (tr).
\item[/capability/imageServiceType (!)]The class of image service: Cutout, Mosaic, Atlas, Pointed (sia).
\item[/capability/interface/securityMethod/@standardID (!)]A description of a security mechanism.  Although this really is a property of an interface, the reference here goes to the embedding capability, which is sufficient for discovery of the existence of such interfaces.  If the capability holds multiple interfaces, clients need other means (e.g., VOSI capabilities) to figure out the mapping between access URLs and supported security methods.
\item[/capability/language/name (!)]A short, human-readable name of a language understood by the TAP service (tr).
\item[/capability/language/version/@ivo-id (!)]The IVORN of a language supported by a TAP service (tr).
\item[/capability/maxAperture]The largest aperture that can be supported upon request via the APERTURE input parameter by a service that supports the special extraction creation method (ssap).
\item[/capability/maxFileSize (!)]The maximum image file size in bytes (sia).
\item[/capability/maxImageExtent/lat]The maximum size in the latitude (Dec.) direction (sia).
\item[/capability/maxImageExtent/long]The maximum size in the longitude (R.A.) direction (sia).
\item[/capability/maxImageSize/lat]The image size in the latitude (Dec.) direction in pixels (sia-1.0).
\item[/capability/maxImageSize/long]The image size in the longitude (R.A.) direction in pixels (sia-1.0).
\item[/capability/maxImageSize]A measure of the largest image the service can produce given as the maximum number of pixels along the first or second axes. (sia).
\item[/capability/maxQueryRegionSize/lat]The maximum size in the latitude (Dec.) direction (sia).
\item[/capability/maxQueryRegionSize/long]The maximum size in the longitude (R.A.) direction (sia).
\item[/capability/maxRecords (!)]The largest number of items (records, rows, etc.) that the service will return (cs, sia, vg, ssap).
\item[/capability/maxSearchRadius (!)]The largest search radius, in degrees, that will be accepted by the service without returning an error condition. Not providing this element or specifying a value of 180 indicates that there is no restriction. (ssap)
\item[/capability/maxSR (!)]The largest search radius of a cone search service (cs).
\item[/capability/outputFormat/@ivo-id (!)]An IVORN of an output format (tr).
\item[/capability/outputFormat/alias]A short, mnemonic identifier for a service's output format (tr).
\item[/capability/outputFormat/mime (!)]The MIME type of an output format (tr).
\item[/capability/outputLimit/default]The maximal output size for newly-created jobs (tr).
\item[/capability/outputLimit/default/@unit]The unit (rows/bytes) in which the service's default output limit is given (tr).
\item[/capability/outputLimit/hard]The hard limit of a service's output size (tr).
\item[/capability/outputLimit/hard/@unit]The unit of this service's hard output limit (tr).
\item[/capability/retentionPeriod/default]The default time between job creation and removal on this service, given in seconds (tr).
\item[/capability/retentionPeriod/hard]The hard limit for the retention time of jobs on this services (tr).
\item[/capability/supportedFrame (!)]The STC name for a world coordinate system frame supported by this service (ssap).
\item[/capability/testQuery/catalog]The catalog to query (cs).
\item[/capability/testQuery/dec]Declination in a test query (cs)
\item[/capability/testQuery/extras]Any extra (non-standard) parameters that must be provided (apart from what is part of base URL given by the accessURL element; cs, sia).
\item[/capability/testQuery/pos/lat]The Declination of the center of the search position in decimal degrees (ssap, sia).
\item[/capability/testQuery/pos/long]The Right Ascension of the center of the search position in decimal degrees (ssap, sia).
\item[/capability/testQuery/pos/refframe]A coordinate system reference frame name for a test query. If not provided, ICRS is assumed (ssap).
\item[/capability/testQuery/queryDataCmd]Fully specified test query formatted as an URL argument list in the syntax specified by the SSA standard. The list must exclude the REQUEST argument (ssap).
\item[/capability/testQuery/ra]Right ascension in a test query (cs).
\item[/capability/testQuery/size]The size of the search radius in an SSA search query (ssap).
\item[/capability/testQuery/size/lat]Region size for a SIA test query in declination (sia).
\item[/capability/testQuery/size/long]Region size for a SIA test query in RA (sia).
\item[/capability/testQuery/sr]Search radius of a cone search service's test query (cs).
\item[/capability/testQuery/verb]Verbosity of a service's test query (cs, sia).
\item[/capability/uploadLimit/default]An advisory size above which user agents should reconfirm uploads to this service (tr).
\item[/capability/uploadLimit/default/@unit]The unit of the limit specified (tr).
\item[/capability/uploadLimit/hard]Hard limit for the size of uploads on this service (tr).
\item[/capability/uploadLimit/hard/@unit]The unit of the limit specified (tr).
\item[/capability/uploadMethod/@ivo-id]The IVORN of an upload method supported by the service (tr).
\item[/capability/verbosity (!)]\texttt{true} if the service supports the VERB keyword; \texttt{false}, otherwise (cs).
\item[/coverage/footprint (!)]A URL of a footprint service for retrieving precise and up-to-date description of coverage (vs).
\item[/coverage/footprint/@ivo-id (!)]The URI form of the IVOA identifier for the service describing the capability refered to by this element (vs).
\item[/deprecated (!)]A sentinel that all versions of the referenced standard are deprecated. The value is a human-readable explanation for the designation (vstd).
\item[/endorsedVersion (!)]A version of a standard that is recommended for use (vstd).
\item[/facility (!)]The observatory or facility used to collect the data contained or managed by this resource (vs).
\item[/format (!)]The physical or digital manifestation of the information supported by a (DataCollection) resource.  MIME types should be used for network-retrievable, digital data, non-MIME type values are used for media that cannot be retrieved over the network (vs).
\item[/format/@isMIMEType]If \texttt{true}, then an accompanying \texttt{/format} item is a MIME Type. Within \rtent{res\_details}, this does not work for services that give more than one format; since furthermore the literal of \vorent{vs:Format} allows a good guess whether or not it is a MIME type, this does not appear a dramatic limitation (vs).
\item[/full]If true, the registry attempts to collect all resource records known to the IVOA (vg).
\item[/instrument (!)]The instrument used to collect the data contained or managed by a resource (vr).
\item[/instrument/@ivo-id (!)]IVORN of the instrument used to collect the data contained or managed by a resource (vr).
\item[/managedAuthority (!)]An authority identifier managed by a registry (vg).
\item[/managingOrg (!)]The organization that manages or owns this authority (vg).
\item[/schema/@namespace (!)]An identifier for a schema described by a standard (vstd).

\end{description}

Note that the representation of StandardsRegExt's 
\vorent{status}  and \vorent{use}
attributes as well as its \vorent{key} would require sequences of
complex objects, which is impossible using \rtent{res\_detail}.
Hence, the respective metadata is not queriable
within the relational registry. Similarly, complex TAPRegExt metadata on
languages, user defined functions, and the like cannot be
represented in this table.  Since these pieces of metadata do not seem
relevant to resource discovery and are geared towards other uses of the
respective VOResource extensions, a more complex model does not seem
warranted just so they can be exposed.



\section{The Extra UDFs in
PL/pgSQL}

\label{appPGDefs}

What follows are (non-normative) 
implementations of three of the user defined functions
specificed in section \ref{adqludf} in the SQL dialect
of PostgreSQL (e.g., \citet{doc:Postgres92}).

Note that PostgreSQL cannot use fulltext indexes on the respective
columns if the functions are defined in this way for (fairly subtle)
reasons connected with NULL value handling.  While workarounds are
conceivable, they come with potentially unwelcome side effects, at least
as of PostgreSQL 9.x.  It is therefore recommended to replace
expressions involving the functions given here with simple boolean
expressions in the ADQL translation layer whenever possible.

\begin{lstlisting}[language=SQL,flexiblecolumns=true]
  CREATE OR REPLACE FUNCTION 
    ivo_hasword(haystack TEXT, needle TEXT)
  RETURNS INTEGER AS $func$
    SELECT CASE WHEN to_tsvector($1) @@ plainto_tsquery($2) 
      THEN 1 
      ELSE 0 
    END
  $func$ LANGUAGE SQL;

  CREATE OR REPLACE FUNCTION 
    ivo_hashlist_has(hashlist TEXT, item TEXT)
  RETURNS INTEGER AS $func$
    -- postgres can't RE-escape a user string; hence, we'll have
    -- to work on the hashlist (this assumes hashlist is already
    -- lowercased).
    SELECT CASE WHEN lower($2) = ANY(string_to_array($1, '#'))
      THEN 1 
      ELSE 0 
    END
  $func$ LANGUAGE SQL;

  CREATE OR REPLACE FUNCTION 
    ivo_nocasematch(value TEXT, pattern TEXT)
  RETURNS INTEGER AS $func$
    SELECT CASE WHEN $1 ILIKE $2
      THEN 1 
      ELSE 0 
    END
  $func$ LANGUAGE SQL;

  -- ivo_string_agg directly corresponds to string_agg; this translation
  -- should be done in the ADQL translator.
\end{lstlisting}



\section{Implementation notes}

\label{appBP}

This appendix contains a set of constraints and recommendations for
implementing the relational registry model on actual RDBMSes, originating
partly from an analysis of the data content of the VO registry in February
2013, partly from a consideration of limits in various XML schema documents.
This concerns, in particular, minimum lengths for columns of type
adql:VARCHAR.  Implementations MUST NOT truncate strings of length equal
to or smaller than the minimal lengths given here; the limitations are
not, however, upper limits, and indeed, when choosing an implementation
strategy it is in general preferable to not impose artificial length
restrictions, in particular if no performance penalty is incurred.

These notes can also be useful with a view to preparing user interfaces for
the relational registry, since input forms and similar user interface
elements invariably have limited space; the limits here give reasonable
defaults for the amount of data that should minimally be manipulatable
by a user with reasonable effort.

The \rtent{ivoid} field present in every table of this
specification merits special consideration, on the one hand due to its
frequency, but also since other IVOA identifiers stored in the
relational registry should probably be treated analogously.
Given that IVORNs in the 2013 data fields have a maximum
length of roughly 100 characters, we propose that a maximum length of 
255 should be sufficient even when taking possible fragment identifiers
into account.

\begin{inlinetable}
\begin{tabular}{llp{6cm}}
\sptablerule
\textbf{Field type}&
\textbf{Datatype}&
\textbf{Pertinent Fields}\\
\sptablerule
ivo-id&\texttt{VARCHAR(255)}&
          {all\_tables}.ivoid\hfil\break
          res\_role.role\_ivoid\hfil\break
          capability.standard\_id\hfil\break
          relationship.related\_id\hfil\break
          validation.validated\_by\hfil\break
          res\_detail.detail\_value for several values of detail\_xpath\\
\sptablerule
\end{tabular}
\end{inlinetable}

The relational registry also contains some date-time values. The most
straightforward implementation certainly is to use SQL timestamps.
Other relational registry fields that straightforwardly map to common
SQL types are those that require numeric values, viz.,
\texttt{REAL}, \texttt{SMALLINT}, and
\texttt{INTEGER}.  The following table summarizes these:

\begin{inlinetable}
\begin{tabular}{llp{6cm}}
\sptablerule
\textbf{Field type}&
\textbf{Datatype}&
\textbf{Pertinent Fields}\\
\sptablerule
floating point&\texttt{REAL}&resource.region\_of\_regard\\
\sptablerule
small integer&\texttt{SMALLINT}&table\_column.std\hfil\break
          intf\_param.std\hfil\break
          validation.level\\
\sptablerule
\end{tabular}
\end{inlinetable}

The fields containing Utypes, UCDs, and Units are treated in parallel
here. The 2013 registry data indicates that a length of 128 characters is
sufficient for real-world purposes; actually, at least UCDs and Units
could of course grow without limitations, but it seems unreasonable
anything longer than a typical line might actually be useful.  As far as
utypes are concerned, we expect those to shrink rather than grow with
new standardization efforts.

In this category, we also summarize our resource xpaths.  With the
conventions laid down in this document, it seems unlikely that future
extensions to VOResource will be so deeply nested that 128 characters
will not be sufficient for their resource xpaths.

\begin{inlinetable}
\begin{tabular}{llp{6cm}}
\sptablerule
\textbf{Field type}&
\textbf{Datatype}&
\textbf{Pertinent Fields}\\
\sptablerule
Utype&\texttt{VARCHAR(128)}&res\_schema.schema\_utype\hfil\break
res\_table.table\_utype\hfil\break
table\_column.utype\hfil\break
intf\_param.utype\\
\sptablerule
UCD&\texttt{VARCHAR(128)}&
          table\_column.ucd\hfil\break
          intf\_param.ucd\\
\sptablerule
Unit&\texttt{VARCHAR(128)}&
          table\_column.unit\hfil\break
          intf\_param.unit\\
\sptablerule
xpath&\texttt{VARCHAR(128)}&
          res\_detail.detail\_xpath \\
\sptablerule
\end{tabular}
\end{inlinetable}

The relational registry further has an
e-mail field, for which we chose 128 characters as a reasonable upper
limit (based on a real maximum of 40 characters in the 2013 data).
There are furthermore URLs (in addition to access and reference URLs,
there are also URLs for the WSDL of SOAP services and logos for roles).
Due to the importance of in particular the access URLs we strongly
recommend to use non-truncating types here.  Empirically, there are
access URLs of up to 224 characters in 2013 (reference URLs are less
critical at a maximum of 96 characters).  Expecting that with REST-based
services, URL lengths will probably rather tend down than up, we still
permit truncation at 255 characters.

\begin{inlinetable}
\begin{tabular}{llp{6cm}}
\sptablerule
\textbf{Field type}&
\textbf{Datatype}&
\textbf{Pertinent Fields}\\
\sptablerule
e-mail&\texttt{VARCHAR(128)}&
          res\_role.email\\
\sptablerule
URLs&\texttt{VARCHAR(255)}&resource.reference\_url\hfil\break
          res\_role.logo\hfil\break
          interface.wsdl\_url\hfil\break
          interface.access\_url \\
\sptablerule
\end{tabular}
\end{inlinetable}

The next group of columns comprises those that have values taken from
a controlled or finite vocabulary.
Trying to simplify the view,
lengths in the form
of powers of two are considered.

\begin{inlinetable}
\begin{tabular}{llp{6cm}}
\sptablerule
\textbf{Field type}&
\textbf{Datatype}&
\textbf{Pertinent Fields}\\
\sptablerule
long enumerations&\texttt{VARCHAR(255)}&resource.content\_level\hfil\break
          resource.content\_type\\
\sptablerule
short enumerations&\texttt{VARCHAR(64)}&
          resource.rights\hfil\break
          resource.waveband\\
\sptablerule
type names&\texttt{VARCHAR(32)}&resource.res\_type\hfil\break
          capability.cap\_type\hfil\break
          res\_table.table\_type\hfil\break
          table\_column.flag\hfil\break
          table\_column.datatype\hfil\break
          table\_column.extended\_schema\hfil\break
          table\_column.extended\_type\hfil\break
          table\_column.type\_system\hfil\break
          interface.result\_type\hfil\break
          intf\_param.datatype\hfil\break
          intf\_param.extended\_schema\hfil\break
          intf\_param.extended\_type\\
\sptablerule
short terms&\texttt{VARCHAR(4)}&interface.query\_type\hfil\break
          interface.url\_use\\
\sptablerule
\end{tabular}
\end{inlinetable}

Finally, there are the fields that actually contain what is
basically free text.
For these, we have made a choice from reasonable powers of two lengths 
considering the actual lengths in the 2013 registry data. 
A special case are fields that either contain natural language text (the
descriptions) or those that have grown without limit historically
(resource.creator\_seq, and, giving in to current abuses discussed above,
res\_role.role\_name).  For all such fields, no upper limit can sensibly
be defined.  However, since certain DBMSes (e.g., MySQL older than
version 5.6) cannot index fields with a TEXT datatype and thus using
VARCHAR may be necessary at least for frequenly-searched fields, we give
the maximal length of the fields in the 2013 registry in parentheses after
the column designations for the TEXT datatype:

\begin{inlinetable}
\begin{tabular}{llp{6cm}}
\sptablerule
\textbf{Field type}&
\textbf{Datatype}&
\textbf{Pertinent Fields}\\
\sptablerule
free string values&\texttt{TEXT}&resource.res\_description
(7801)\hfil\break
          resource.creator\_seq (712)\hfil\break
          res\_role.role\_name (712)\hfil\break
          res\_schema.schema\_description (934)\hfil\break
          res\_table.table\_description (934)\hfil\break
          table\_column.description (3735)\hfil\break
          intf\_param.description (347)\hfil\break
          capability.cap\_description (100)\\
\sptablerule
titles, etc.&\texttt{VARCHAR(255)}&resource.res\_title\hfil\break
          res\_role.address\hfil\break
          res\_schema.schema\_title\hfil\break
          res\_table.table\_title\hfil\break
          relationship.related\_name\hfil\break
          res\_detail.detail\_value\\
\sptablerule
expressions&\texttt{VARCHAR(128)}&resource.version\hfil\break
          resource.source\_value\hfil\break
          res\_subject.res\_subject\\
\sptablerule
names&\texttt{VARCHAR(64)}&res\_table.table\_name\hfil\break
          table\_column.name\hfil\break
          intf\_param.name\\
\sptablerule
misc.~short strings&\texttt{VARCHAR(32)}&resource.source\_format\hfil\break
          res\_role.telephone\hfil\break
          res\_schema.schema\_name\hfil\break
          interface.intf\_type\hfil\break
          interface.intf\_role\hfil\break
          relationship\_type\hfil\break
          res\_date.value\_role\\
\sptablerule
misc.~particles&\texttt{VARCHAR(16)}&resource.short\_name\hfil\break
          table\_column.arraysize\hfil\break
          intf\_param.arraysize\hfil\break
          interface.std\_version\hfil\break
          intf\_param.use\\
\end{tabular}
\end{inlinetable}



\section{XSLT to enumerate Relational
Registry XPaths}

\label{appMkut}

The (non-normative) following XSL stylesheet emits xpaths as
discussed in section \ref{vorutypes} when applied to a
VOResource extension.  Considering readability and limitations of XSLT, this is
not fully general -- if VOResource extensions started to inherit from other
extensions' subclasses of Resource, Capability, or Interface, this stylesheet
will need to be extended.

Still, it is a useful tool when evaluating how to map a given extension
to the relational registry.

\begin{lstlisting}[language=XML,flexiblecolumns=true]
<?xml version="1.0" encoding="utf-8"?>
<stylesheet version="1.0" 
  xmlns="http://www.w3.org/1999/XSL/Transform"
  xmlns:xs="http://www.w3.org/2001/XMLSchema" 
  xmlns:vr="http://www.ivoa.net/xml/VOResource/v1.0" 
  xmlns:vs="http://www.ivoa.net/xml/VODataService/v1.1">

<!-- extract RegTAP xpath from VOResource and related XML schema files.
This file is in the public domain. -->

<!-- The basic strategy here is: start from all discernible types
derived from vr:Resource, vr:Capability, and vr:Interface; we need to
handle capability and interface separately since they may not be
explicitely reachable from a resource due to the way capability and
interface types are declared in VOResource (i.e., through xsi:type).

Each root is the start element for a recursion yielding the utypes.
During this recursion, attributes yield utypes by concatenating the
parent name with the attribute name, and elements yield utypes by
concatenating the element name with the parent name.  Elements take part
in the recursion, where their utype becomes the new parent name.

Note that the xpaths generated here are the ones used in the 
res_details table.  For the xpaths used in lieu of utypes in
TAP_SCHEMA and VOSI tables, you will want to make them relative.

Hack alert: We need to traverse the type tree; however, due to 
(practical) limitations of XSLT1, we do not do that across files.  So,
if a type were to inherit from a class derived from VOResource or Capability
in another document, this stylesheet would not notice.  Also, we kill all
namespace prefixes in attributes.  Proper handling of that probably is
close to impossible with XSLT1.  -->

<output method="text"/>
<strip-space elements="*"/>

<!-- An index used to retrieve the type definitions for the child
elements in walk; note that this only spans one file,
which is the primary limitation of this program -->
<key name="definitions" match="xs:simpleType|xs:complexType"
  use="@name"/>

<!-- A map from types to the names of their base classes; this is
a bit haphazard since we don not want to use XSLT extensions that
actually understand XML schema.  A recursion within the type
hierarchy takes place in walk-for-type -->
<key name="of-type" match="xs:complexType"
  use="substring-after(./xs:complexContent/*/@base, ':')"/>

<template name="add-doc">
  <!-- retrieve the "short" doc and add them in parens if present,
  emit an lf either way -->
  <if test="boolean(xs:annotation[1])">
     <value-of select="concat(' (', normalize-space(xs:annotation[1]/xs:documentation), ')')"/>
  </if>
   <text>&#10;</text>
</template>

<template name="walk">
  <!-- the central (recursive) template building up utypes by
  collecting subelements; the current node here is a type definition -->

  <param name="parent-path"/>

  <for-each select="descendant::xs:attribute">
    <value-of select="concat($parent-path, '/@', @name)"/>
    <call-template name="add-doc"/>
  </for-each>

  <for-each select="descendant::xs:element">
    <!-- capability and interface are roots of their own -->
    <if test="@name!='capability' and @name!='interface'">
      <value-of select="concat($parent-path, '/', @name)"/>
      <call-template name="add-doc"/>

      <variable name="child-type"
        select="substring-after(@type, ':')"/>
      <if test="key('definitions', $child-type)">
          <variable name="child-path"
            select="concat($parent-path, '/', @name)"/>
        <for-each select="key('definitions', $child-type)">
          <call-template name="walk">
            <with-param name="parent-path" select="$child-path"/>
          </call-template>
        </for-each>
      </if>
    </if>
  </for-each>
 
  <!-- collect attributes and elements for our node from the types
  we are derived from, too -->
  <for-each select="descendant::xs:extension|descendant::xs:restriction">
    <for-each select="key('definitions', substring-after(@base, ':'))">
      <call-template name="walk">
        <with-param name="parent-path" select="$parent-path"/>
      </call-template>
    </for-each>
  </for-each>
</template>

<template name="walk-for-type">
  <!-- iterates over the types and elements derived from base-type -->
  <param name="base-type"/>
  <param name="parent-path"/>
  <for-each select="key('of-type', $base-type)">

    <call-template name="walk">
      <with-param name="parent-path" select="$parent-path"/>
    </call-template>

    <call-template name="walk-for-type">
      <with-param name="base-type" select="@name"/>
      <with-param name="parent-path" select="$parent-path"/>
    </call-template>

  </for-each>
</template>

<!-- templates cover the immediate types; the derived types are handled
in explicit call-templates in the the template for root below. -->
<template match="//xs:complexType[@name='Resource']">
  <call-template name="walk">
    <with-param name="parent-path" select="'Resource'"/>
  </call-template>
</template>

<template match="//xs:complexType[@name='Capability']">
  <call-template name="walk">
    <with-param name="parent-path" select="'Capability'"/>
  </call-template>
</template>

<template match="//xs:complexType[@name='Interface']">
  <call-template name="walk">
    <with-param name="parent-path" select="'Interface'"/>
  </call-template>
</template>

<template match="/">
  <apply-templates/>

  <call-template name="walk-for-type">
    <with-param name="base-type" select="'Resource'"/>
    <with-param name="parent-path" select="''"/>
  </call-template>

  <call-template name="walk-for-type">
    <with-param name="base-type" select="'Service'"/>
    <with-param name="parent-path" select="''"/>
  </call-template>

  <call-template name="walk-for-type">
    <with-param name="base-type" select="'Capability'"/>
    <with-param name="parent-path" select="'/capability'"/>
  </call-template>

  <call-template name="walk-for-type">
    <with-param name="base-type" select="'Interface'"/>
    <with-param name="parent-path" select="'/capability/interface'"/>
  </call-template>

</template>

<template match="text()"/> 

</stylesheet>
\end{lstlisting}



\section{Changes from Previous Versions}

\label{changes}

\subsection{Changes from PR-2014-10-30}

\begin{itemize}
\item No changes to specification content (only minor typo fixes).
\end{itemize}


\subsection{Changes from PR-20140627}

\begin{itemize}
\item Removed reference to a future STC extension.
\item Migrated to ivoatex.
\end{itemize}

\subsection{Changes from PR-20140227}

\label{changes-20140227}

\begin{itemize}

\item Added a \texttt{/full} details xpath from VORegistry (this had
  been forgotten due to limitations in the makeutypes stylesheet).{}

\item Added a \texttt{/capability/interface/securityMethod/@standardID}
  details xpath from vr:Interface.{}

\item Added requirement to implement the \rtent{ivo\_string\_agg}
  user defined function.{}

\item Added a section specifying the treatment of non-ASCII characters
  in RegTAP columns.{}

\item New rules on string normalization: strings must be
  whitespace-stripped, empty strings must be mapped to NULL.{}

\item Dropped requirements that the \texttt{\_index} columns are
  integers (let alone small integers); added a section discussing in
  what sense they are implementation defined.{}

\item Dropped adql: prefixes on TAP\_SCHEMA.columns datatypes.{}

\item Now declaring a precedence of xpaths generated by rules over the
  list in Appendix \ref{d_u_list}.{}

\item Clarified translation of column/@std and param/@std.{}

\item Now recommending to constrain on \rtent{intf\_type}
  (rather than \rtent{intf\_role}, as before) when locating standard
  interfaces.{}

\item Redactional changes from RFC (e.g., in column descriptions, some 
  clarifications, typo fixes).{}

\end{itemize}



\subsection{Changes from WD-20131203}

\label{changes-20131203}

\begin{itemize}

\item Changed the data model identifier to
  \texttt{ivo://ivoa.net/std/RegTAP\#1.0} to match usage with a
  later standards record.{}

\item Fixed a typo in a column name: schema.schemaname is now schema.schema\_name
  as in the prose.{}

\item Recovered 
  \texttt{/capability/uploadMethod/@ivo-id} res\_detail keys that was
  accidentally lost in a previous version.{}

\item Clarification of nomenclature.{}

\end{itemize}



\subsection{Changes from WD-20130909}

\label{changes-20130909}

\begin{itemize}

\item Updates for REC of SimpleDALRegExt, which contains versions 1.1 of
  both the sia and the ssap XML schemas; this means there are now additional
  namespace URIs to take into accound, as well as new res\_detail xpaths
  \texttt{/capability/maxSearchRadius}, 
  \texttt{/capability/maxImageSize}, and 
  \texttt{/capability/testQuery/pos/refframe}.{}

\item Reinstated makeutypes.xslt script; it's useful even with the new
  xpaths.{}

\end{itemize}



\subsection{Changes from WD-20130411}

\label{changes-20130411}

\begin{itemize}

\item The final utype reform: most of our ex-utype strings aren't called utypes
  any more, they're fairly plain xpaths.  Consequently, 
  \rtent{res\_detail.detail\_utype} has been renamed 
  \rtent{res\_detail.detail\_xpath}.{}

\item Renamed some columns and the subject table to relieve the need of quoting
  in MS SQL Server (or, in the case or \rtent{use\_param}, maintain
  consistency after the renaming):\\

\begin{tabular}{lll}

\textbf{Old}&
\textbf{New}\\
resource.version&resource.res\_version\\
res\_role.address&res\_role.street\_address\\
subject.*&res\_subject.*\\
res\_subject.res\_subject&res\_subject.res\_subject\\
table\_column.description&table\_column.column\_description\\
intf\_param.description&intf\_param.param\_description\\
intf\_param.use\_param&intf\_param.param\_use\\
validation.level&validation.val\_level\\

\end{tabular}

\item rr.intf\_param grew the arraysize and delim columns that before
    accidentally were only present in rr.table\_column.{}

\item Added warnings about having to match case-insensitively in
  res\_detail.detail\_value for IVORN-valued rows.{}

\item Restored the foreign key from interface to capability.  Mandating
  ignoring interface elements from StandardsRegExt records really is the
  lesser evil.{}

\item \rtent{resource.region\_of\_regard} now must have unit metadata
  declared.{}

\item We now explicitely deprecate multiple access URLs per interface
  and announce that single access URLs will be enforced in future
  VOResource versions.{}

\end{itemize}



\subsection{Changes from WD-20130305}

\label{changes-20130305}

Changes from WD-20130305
\begin{itemize}

\item intf\_index is now required to be unique within a resource, not a
capability; this is because StandardsRegExt has interfaces outside
of capabilities.  In consequence, the intf\_param no longer has a
cap\_index column, and its foreign key is just ivoid and intf\_index.{}

\item Added handling for the StandardsRegExt schema element.{}

\item The list of res\_details utypes was moved to an appendix since
it was too long to be included in the running text.{}

\item Redation for WD publication.{}

\end{itemize}



\subsection{Changes from WD-20121112}

\label{changes-20121112}

Changes from WD-20121112
\begin{itemize}

\item Adapted all utypes to better match future VO-DML utypes.{}

\item footprint, data\_url, facility, and instrument are no longer in rr.resource
but are instead kept in rr.res\_details rows.{}

\item For VOResource compliance, intf\_param has no flag column any more.{}

\item res\_role.base\_utype is renamed to res\_role.base\_role and no longer
pretends to be a utype fragment; also, the content is now a simple
word..{}

\item intf\_param.use is now called intf\_param.use\_param to avoid possible
clashes with reserved SQL words.{}

\item Removed all material on STC coverage.{}

\item Added an appendix recommending field sizes.{}

\end{itemize}




\bibliography{ivoatex/ivoabib,local}

\end{document}